\documentclass[nofootinbib,superscriptaddress,twocolumn]{revtex4}
\usepackage{graphicx}
\usepackage[english]{babel}
\usepackage{amsmath,amssymb}

\newcommand{\ie}{\emph{i.e., }}
\newcommand{\reff}[1]{(\ref{#1})}
\newcommand{\eref}[1]{Eq.\reff{#1}}
\newcommand{\erefs}[1]{Eqs.\reff{#1}}
\newcommand{\figref}[1]{FIG.\ref{#1}}
\newcommand{\citer}[1]{Ref.\cite{#1}}
\newcommand{\citers}[1]{Refs.\cite{#1}}
\newcommand{\omp}{\omega_p}
\newcommand{\p}{\partial}

\usepackage{color}

\begin{document}
\title{One dimensional reduced model for ITER relevant energetic particle transport}

\author{N. Carlevaro}
\affiliation{ENEA, Fusion and Nuclear Safety Department, C. R. Frascati,\\
             Via E. Fermi 45, 00044 Frascati (Roma), Italy}
\affiliation{CREATE Consortium, Via Claudio 21 (80125) Napoli, Italy}

\author{G. Meng}
\affiliation{Max Planck Institute for Plasma Physics, Boltzmannstrasse 2, D-85748
             Garching, Germany}
             
\author{G. Montani}
\affiliation{ENEA, Fusion and Nuclear Safety Department, C. R. Frascati,\\
             Via E. Fermi 45, 00044 Frascati (Roma), Italy}       
\affiliation{Physics Department, ``Sapienza'' University of Rome,\\
             P.le Aldo Moro 5, 00185 Roma (Italy)}

\author{F. Zonca}
\affiliation{ENEA, Fusion and Nuclear Safety Department, C. R. Frascati,\\
             Via E. Fermi 45, 00044 Frascati (Roma), Italy}
\affiliation{Institute for Fusion Theory and Simulation and Department of Physics,\\
             Zhejiang University, Hangzhou 310027, China}

\author{T. Hayward-Schneider}
\affiliation{Max Planck Institute for Plasma Physics, Boltzmannstrasse 2, D-85748
             Garching, Germany}

\author{Ph. Lauber}
\affiliation{Max Planck Institute for Plasma Physics, Boltzmannstrasse 2, D-85748
             Garching, Germany}

\author{Z. Lu}
\affiliation{Max Planck Institute for Plasma Physics, Boltzmannstrasse 2, D-85748
             Garching, Germany}

\author{X. Wang}
\affiliation{Max Planck Institute for Plasma Physics, Boltzmannstrasse 2, D-85748
             Garching, Germany}


\begin{abstract}
We set up a mapping procedure able to translate the evolution of the radial profile of fast ions, interacting with Toroidal Alfv\'en Eigenmodes, into the dynamics of an equivalent one dimensional bump-on-tail system. We apply this mapping technique to reproduce ITER relevant simulations, which clearly outlined deviations from the diffusive quasi-linear (QL) model. Our analysis demonstrates the capability of the one-dimensional beam-plasma dynamics to predict the relevant features of the non-linear hybrid LIGKA/HAGIS simulations. In particular, we clearly identify how the deviation from the QL evolutive profiles is due to the presence of avalanche processes. A detailed analysis regarding the reduced dimensionality is also addressed, by means of phase-space slicing based on constants of motion. In the conclusions, we outline the main criticalities and outcomes of the procedure, which must be satisfactorily addressed to make quantitative prediction on the observed outgoing fluxes in a Tokamak device.
\end{abstract}
\maketitle

\section{Introduction}
One of the key issues for magnetic confinement fusion is the excitation of global instabilities by energetic particles (EPs) in burning plasmas. The related non-linear EP transport processes (in particular, the redistribution from the core to the edge) are crucial for the plasma heating and for the safety of the Tokamak machine.  The capability of current numerical approaches (such as global magneto-hydrodynamics hybrid-kinetic or fully gyro-kinetic codes) to properly address the non-linear saturation of EP-driven modes in present-day and ITER-size devices is well established. However, such an approach is extremely costly and time consuming, and typically covers only few milliseconds of dynamics or rely on the separation of various time scales capturing different physical processes within distinct phases of the simulation \citep{bierwage18,La13}. Moreover, these phenomena are intrinsically multi scale processes \cite{ZCrmp,ZC15njp,dz13,bass10}, since EPs play as mediators of cross scale couplings \cite{ZC15ppcf}.

For routine applications, EP profile evolution is frequently evaluated by means of reduced transport EP models, such as critical gradient \cite{waltz15}, quasi-linear (QL) \cite{gore18} or the kick-model \cite{pode14}. 
In particular, EP relaxation processes are typically addressed invoking the marginal stability assumption and the stiffness of the distribution \cite{angioni09,Heidbrink13,todo16} (see also \citer{waltz14}). Moreover, test particle methods \cite{hsu92a,hsu92b,FQ13} are generally used when facing shear Alfv\'en wave fluctuations and lower frequency magneto-hydrodynamics modes. 
As already metioned, in the case of multiple discrete Alfv\'en eigenmodes (AEs) with overlapping resonances, the typically adopted reduced description is based on the QL models \cite{BB95b,BB96b}. Recent applications and extensions of this approach to DIII-D scenarios can be found in \citers{GG12,GBG14}, while the so-called Trapped-Gyro-Landau-Fluid model is tested in \citer{sheng17}. Validation analyses of the 1.5D critical gradient model \cite{GG12,waltz15} for EP redistribution are instead discussed in \citer{gore16}. This actually corresponds to the marginal stability limit of the QL approach and relies on the linear stability analysis of AEs. 
Still requiring for a reduction of the computational effort, in \citer{bourdelle07} a QL gyrokinetic model (QuaLiKiz) is developed by addressing the leading order local fluctuation structure and particle response as further simplification. See \citer{bourdelle16}, for a detailed comparison between fully non-linear and reduced schemes for turbulence driven EP transport. While QL models are predictive in facing EP redistribution in specific cases \cite{ZCrmp}, in \citers{white10a,white10b} it is pointed out how realistic predictions for stochastization threshold and transport in the ITER scenario are strongly dependent on the underlying physics details and on the phase space characteristics of the EP distribution \cite{pode14,pode16}. Moreover, according to fully non-linear hybrid magnetohydrodynamics-gyrokinetic  studies \cite{ZC15ppcf,ZCrmp,vlad18}, both phase bunching and locking deeply affect EP transport, and they are not accounted for in QL reduced descriptions \cite{ZC15njp}. Meanwhile, the linear stable part of the fluctuating spectrum is found to play a crucial role in the EP dynamics \cite{spb16,ncentropy} (see also \citers{meng18,zacha16}).

Among reduced schemes, also the 1D bump-on-tail (BoT) paradigm has been addressed to describe the non-linear interactions between Alfv\'enic fluctuations and EPs \cite{BB90a,BB90b,BB90c,CZ13,ZC15njp}. In \citers{BS11,ZCrmp}, a comprehensive analysis of the applications and restrictions of the BoT model is given. See also \citer{nceps16}, for a comparison with respect to the single beta-induced AE non-linear dynamics \cite{xin16}. The 1D model, as a powerful theoretical and numerical tool, has also been adopted to demonstrate the essence of the optimization of the implicit particle scheme in Tokamak geometry \cite{zl21}.

In this paper, with the aim of finding and validating reduced models able to reproduce the relevant features of the AE/EP interaction, we define a mapping procedure between the radial coordinate of the realistic 3D scenario and the velocity dimension of the BoT paradigm (described by an Hamiltonian beam-plasma model \cite{CFMZJPP}). Although the BoT model has been widely investigated in the literature, the present analysis has the merit to address a quantitative treatment of the EP profile relaxation as predicted by a 1D model in comparison to LIGKA/HAGIS simulations of a realistic ITER scenario. The mapping consists in a one-to-one correspondence between the velocity space and radial variable and it is defined around a single reference resonance, resulting in a linear relation, then extended to the multiple mode case outlining the critical issues and advantages of the addressed procedure. The BoT setup is closed by determining the beam density through the linear analysis. The density parameter is defined imposing a linear scaling law between the instability drive (normalized to the mode frequency) in the two approaches. The resulting proportionality constant is fixed to optimize proper non-linear EP spread. In fact, it is worthwhile recalling that the EP response of the two systems is intrinsically different. Primarily, the difference stands in the different dimensionality and corresponding equilibrium geometry. Furthermore, radial mode structures are of crucial importance for EP radial transport in fusion plasmas \cite{ZCrmp,ZC15ppcf,ZC15njp}, while such effect is completely absent on the EP redistribution in velocity space in the BoT case. The present reduced model approach has the advantage that, in order to be set up, it only requires information about linear physics from the corresponding LIGKA/HAGIS simulation of the considered realistic scenario. 

The map is applied to the ITER 15MA beseline scenario analyzed in \citer{spb16}, in the presence of the least damped 27 toroidal AEs (TAEs). The drive of the BoT scheme is lowered up to $60\%$ in order to compare the two mapped systems, while the (normalized) dampings are preserved to maintain the correct asymptotic mode decays. The motivation of this choice is, again, connected with the different dimensionality of the two models and will be further discussed and motivated in the following. By means of numerical simulations of the introduced Hamiltonian BoT model, the mode saturation and EP redistribution are analyzed with respect to the results presented in \citer{spb16}. In particular, avalanche effects on the spectral evolution and \emph{domino} (convective) transport well emerges from simulations. This illuminates the importance of the stable (sub dominant) spectrum component for the EP dynamics. Discrepancy and agreement with respect to the effective EP redistribution of \citer{spb16} is analyzed, together with the importance of
the global mode structure due to the alignment of multiple AE gaps leading to a broad poloidal mode number spectrum.

As in \citer{spb16}, particular attention is given to the comparison of the fully self-consistent results with respect to QL predictions. QL equations for the BoT paradigm are mapped to the radial coordinate and numerically integrated for the addressed multi mode case. The obtained QL profiles, according to \citer{spb16}, differ from non-linear simulations since they clearly do not reproduce avalanche phenomena and outer redistribution of EPs.

Regarding the linear scaling law between the instability drives assumed in the mapping procedure, we provide the details of such scaling by analyzing the evolution of a given set of markers which are defined in order to maximize the wave-particle power exchange. The chosen EP marker set, consisting of resonant particles, has the same power transfer as the resonant EP particles in the chosen ITER reference case. However, due to the difference in the equilibrium geometry and in the number of degrees of freedom, the fraction of resonant particles cannot be the same in the two cases. This crucial element will be taken into account by properly reducing the drive in the 1D system. 
This analysis sheds light on the limitations of adopting a reduced model with one degree of freedom for the description of EP behaviors in higher dimensionality. It also provides the seed for further developments in view of a detailed quantitative prediction of EP redistribution by means of 1D models.

The paper is organized as follows. In Sec.\ref{sec_bot}, the Hamiltonian formulation of the BoT paradigm is addressed specifying the normalization adopted in the simulation. In Sec.\ref{sec_map}, the map between the radial coordinate of the AE/EP system and the velocity space of the BoT model is described. Here, critical issues and advantages of such a local procedure are pointed out for the case of multiple resonances. All the ingredients required for the beam plasma non-linear simulations are then determined. By means of this map, the QL model is also derived for the radial EP transport. In Sec.\ref{sec_sims}, the map is applied to the ITER 15MA beseline scenario discussed in \citer{spb16} in the presence of 27 TAE resonances. After the parameter setup and the analysis of the linear features of the two systems, the non-linear simulation results are compared by discussing the spectral evolution and EP transport (avalanche redistribution). Discrepancies with and shortcomings of the QL scheme are also pointed out. 
In in Sec.\ref{alpha_sec}, an analysis of the redistribution of a given set of EPs maximizing the power exchange is provided, outlining the very good predictive nature of the reduced model when the drive is matched with the realistic case. Concluding remarks follows.

\section{Hamiltonian description of the bump-on-tail model}\label{sec_bot}
The BoT problem address a fast electron beam interacting with a 1D plasma considered as a cold dielectric medium, which supports longitudinal electrostatic Langmuir waves (the background plasma density $n_p$ is taken much larger than that of the beam $n_B$) \cite{OM68,OWM71,VK12,EEbook}. In this paper, the BoT dynamics is described using the Hamiltonian formulation addressed in \citers{CFMZJPP,ncentropy} and refs. therein. In particular, the broad EP beam self-consistently evolves in the presence of $M$ modes taken at the plasma frequency: $\omega_j\simeq\omp$, where $j=1,\,...,\,M$. In this case, the (cold) plasma dielectric function \ie $\epsilon=1-\omp^2/\omega_j^2$, is nearly vanishing and a mode is resonantly driven (linearly unstable) when its wave number $k_j$ obeys the resonance rule $k_j=\omega_j/v_{rj}\simeq\omp/v_{rj}$, where $v_{rj}$ denotes a given resonant velocity. The evolutive equations of the BoT model finally reads
\begin{equation}\label{mainsys1}
\begin{split}
&\bar{x}_i'=u_i \;,\\
&u_i'=\sum_{j=1}^{M}\big(i\,\ell_j\;\bar{\phi}_j\;e^{i\ell_j\bar{x}_{i}}+c.c.\big)\;,\\[-8pt]
&\bar{\phi}_j'=-i\bar{\phi}_j-\bar{\gamma}_{dj}\bar{\phi}_j+\frac{i\eta}{2\ell_j^2 N}\sum_{i=1}^{N} e^{-i\ell_j\bar{x}_{i}}\;,
\end{split}
\end{equation}
where we introduced the parameter $\eta=n_B/n_p\ll1$ and the notation is as follows. Beam particle trajectories are normalized using $\bar{x}_i=x_i(2\pi/L)$, where $x_i$ (with $i=1,\,...,\,N$) denotes the beam particle position in the 1D plasma slab of periodicity $L$. The Langmuir potential $\varphi(x,t)$ is described using its Fourier components $\varphi_j(k_j,t)$. Other quantities are normalized as: $\tau=t\omp$, $\ell_j=k_j(2\pi/L)^{-1}$ (integers), $u_i=\bar{x}_i'=v_i(2\pi/L)/\omp$, $\phi_j=(2\pi/L)^2 e\varphi_j/m\omp^2$ and $\bar{\phi}_j=\phi_j e^{-i\tau}$. The prime represents $\tau$ derivative while barred frequencies (growth rates) are in $\omp$ units: $\bar{\omega}_j=\omega_j/\omp$ ($\bar{\gamma}_j=\gamma_j/\omp$). 

We include an external damping $\bar{\gamma}_{dj}$ for each mode, so the growth rate results $\bar{\gamma}_j=\bar{\gamma}_{Dj}-\bar{\gamma}_{dj}$ (where $\bar{\gamma}_{D}$ denotes the normalized linear drive). We also stress how the resonance conditions are set as $\ell_j u_{rj}\simeq1$.

In the non-linear simulations of this paper, \erefs{mainsys1} are solved using a Runge-Kutta (fourth order) algorithm. The broad initial beam is initialized in the velocity space according to a given distribution $F_{B0}(u)$ (for a total of $N=10^{6}$ particles). The details on the distribution function will be discussed in the next Sections. Particle positions $\bar{x}_i(0)$ are given uniform between $0$ and $2\pi$ and the modes are initialized of the order $\mathcal{O}(10^{-14})$, thus the linear regime is guaranteed \emph{ab initio}.

\section{Mapping procedure}\label{sec_map}
Here we describe a mapping procedure between the radial coordinate $r$ of EPs interacting with the Alfv\'enic spectrum and the BoT velocity ($v$) space. This is a one-to-one link between the two corresponding independent variables and then it will be applied to the case presented in \citer{spb16} treating the ITER 15MA baseline scenario specified for the most relevant TAEs.

For a chosen single reference resonance, the mapping is formally derived by means of the resonance condition. Following \citer{ZCrmp,ZC15njp,BW14pop}, in the presence of EPs we can define the resonant frequency $\Omega_F$ as
\begin{align}
\Omega_F(r)\equiv n\omega_D+((n\bar{q}-m)+k_b)\omega_b\;,
\end{align}  
for circulating particle, or 
\begin{align}
\Omega_F(r)\equiv n\omega_D+k_b\omega_b\;,
\end{align}
for trapped EP. Here $n$ denotes the toroidal mode number ($m$ the poloidal one), $\omega_D$ and $\omega_b$ are the precession and bounce/transit frequencies, respectively, while $k_b$ indicates the bounce/transit harmonics and $\bar{q}$ is the orbit integral of the safety factor. Furthermore, $r$ denotes a radial like invariant of motion \cite{ZCrmp,ZC15njp} that, hereafter, will be mapped into the radial coordinate $r$. 

For an AE of frequency $\Omega$, the resonance condition can be written as $\Omega-\Omega_F(r)=0$ and, using two suitable normalization constants $A_{1,2}$, the map can be formally written as
\begin{align}
\frac{\Omega_F(r)-\Omega}{A_1}=\frac{k(v-v_r)}{A_2}\;,
\end{align}
where, as already indicated, the subscript $r$ denotes the resonant value of a quantity. Moreover, for the realistic 3D scenario, we will use capital letters and standard normalization for convenience of notation: the normalized Tokamak radius reads $s=r/a$ ($a$ denotes the minor radius), while, barred frequencies (growth rates and damping) are $\bar{\Omega}=\Omega/A_1$ ($\bar{\Gamma}=\Gamma/A_1$) with $A_1=\omega_{A0}$, where $\omega_{A0}=v_{A0}/R_0$ and $v_{A0}$ is the Alfv\'en speed at the magnetic axis while $R_0$ the major radius.

In this work, we address a local map through an expansion of $\bar{\Omega}_F$ near $s_{r}$ (the resonant normalized radius), where we obviously have $\bar{\Omega}_F(s_r)=\bar{\Omega}$. This assumption is a natural prescription as far as the resonance frequency is sufficiently smooth in $s$ over the considered resonance region. Actually, the narrow character of the TAE spectrum that will be analyzed in the next Sections guarantees that this is a good and reasonable choice. Specifically, we use the linear expansion $\bar{\Omega}_F(s)=\bar{\Omega}+(s-s_r)\p_s\bar{\Omega}_F$. Due to the inverse Landau damping mechanism at resonance, $A_2$ must have the opposite sign of $\p_s\bar{\Omega}_F$ in order to obtain the correct slope of the distribution function in the velocity space. Thus,
\begin{align}\label{fkablksdbvlhj}
v=v_r-\frac{|A_2 \p_s\bar{\Omega}_F|}{k}\,(s-s_r)\;.
\end{align}
Imposing the boundary condition $s=0\;\mapsto\;v=v_{max}$ and $s=1\;\mapsto\;v=0$, we get $v_{max}=|A_2\p_s\bar{\Omega}_F|/k$ and $v_r=v_{max}(1-s_r)$. For this part of the mapping procedure, dealing with fixed frequency modes supported by the background plasma, we can write the resonance condition for the BoT paradigm as $v_r k=\omp$. This yields to the following relation:
\begin{align}
|A_2\p_s\bar{\Omega}_F|/\omp=(1-s_r)^{-1}\equiv C_1\;.
\end{align}
In this scheme, we can finally get
\begin{align}\label{mapvs}
v=C_1\omega_p(1-s)/k\;.
\end{align}
In order to fix the resonant wave number $k$ of the reference mode, we also consider the condition $k^*v_{max}=\omp$ (with $v_{max}=C_1v_r$) which define the spectral features (the periodicity length) without loss of generality. This yields to $k=C_1 k^*$, determined by arbitrarily fixing $k^*$. Rewriting the map using the normalized variable for the BoT model described in the previous Section, \eref{mapvs} reads
\begin{align}
u=(1-s)/\ell^*\;.	
\end{align}

This completely defines the map between the velocity and the radial coordinate in terms of dimensionless variables. Meanwhile, BoT system is well defined once the density parameter $\eta$ is fixed in \erefs{mainsys1} and this can be done using the linear dispersion relation \cite{OM68,LP81}. Since the dielectric has been expanded near $\omega\simeq\omp$ in the governing equations, the dispersion relation must be evaluated for $\epsilon\simeq2(\bar{\omega}-1)$, accordingly. The instability features of a single resonant mode, can be analyzed by explicitly expressing $\bar{\omega}\to \bar{\omega}+i\bar{\gamma}_D$, where now $\bar{\omega}$ can include a real frequency shift with respect to the Langmuir mode frequency $\omp$ and we recall that $\bar{\gamma}_D$ is the linear drive. We can write the normalized dispersion relation for a give mode as \cite{jpp_cmms}
\begin{align}\label{disrel}
2(\bar{\omega}+i\bar{\gamma}_D-1)-\frac{\eta}{M\ell}
\int_{-\infty}^{+\infty}\!\!\!\!\!\!\!du\frac{\p_u F_{B0}(u)}{u\ell-\bar{\omega}-i\bar{\gamma}_D}=0\;,
\end{align}
where $M=\int du F_{B0}$. As initial profile, we set the typical slowing down EP profile assumed in the realistic scenario. This can be described by the following distribution function expressed in the radial coordinate:
\begin{align}\label{fhot}
F_{H0}(s)=B_1\,\textrm{Erfc}[B_2+s B_3]\;,
\end{align}
where $B_{1,2,3}$ are given constants. The initial condition for BoT simulations are thus given by $F_{B0}(u)=F_{H0}(1-\ell_1 u)$.

At this point, it is important to stress how the EP/AE system has an higher dimensionality (3D) with respect to the BoT paradigm (1D). In particular, the radial profile is obtained as velocity space average over resonant as well as non-resonant particles, and the global radial EP redistribution takes into account all the corresponding wave-particle energy exchange. Meanwhile, the 1D transport is determined by only resonant particle that maximize the power transfer, and the non-linear transport in the BoT scheme results more efficient than radial transport in the realistic scenario, for equal drive. Thus the following scaling between linear drives normalized to the effective mode frequency is expected and assumed to properly compare the non-linear evolution of the two systems:
\begin{align}\label{scaling}
\bar{\gamma}_D/\bar{\omega}=\alpha\;\bar{\Gamma}_D/\bar{\Omega}\;,\qquad \alpha\leqslant1\;,
\end{align}
(the last condition ensures the reduction of the growth rate in the 1D simulations). We will come back to this point in Sec.\ref{alpha_sec}, where we show that $\alpha$ approaches 1 when comparing only EPs which maximize the energy exchange in the realistic scenario. We will also trace, as a future work perspective, the methodology for estimating the parameter $\alpha$ by invoking the splitting of the phase space due to the proper definition of the motion constants. We stress here that, for the scope of the present analysis, in the numerical study of Sec.\ref{sec_sims} the scaling parameter is instead derived by a comparison of the flattening of the relaxed profiles in the two schemes (see the following Section for further details). In order to preserve the temporal asymptotic mode decay after saturation, we instead assume to retain the same normalized damping for the two mapped systems, \ie $\bar{\gamma}_d/\bar{\omega}=\bar{\Gamma}_d/\bar{\Omega}$. This is consistent with the fact that damping is provided by the background plasma and not by the resonant EPs. In particular, it has been shown in \citer{pinches2005} that, for the ITER scenario analyzed in this paper, the neutral-beam injection (NBI) contribution is destabilizing due to its high beam energy (1 MeV). Thus the damping comes from the background species only in the form of mainly Landau damping and radiative damping. In \citer{spb16} the NBI drive was not included (thus not relevant for this analysis), however a scan in the alpha particle drive was carried out in order to demonstrate how the non-linear evolution depends on additional EP drive.

The linear dispersion relation \eref{disrel} can be now numerically integrated providing the values of the mode frequency $\bar{\omega}$ and the parameter $\eta$. This formally closes the mapping procedure.

Let us now discuss how to add multiple modes in the dynamics. In such a case, the mapping paradigm should be necessarily specified for one single reference resonance and then, once the parameter setup is completed, the whole spectrum should be addressed by means of the proper resonance conditions: in this way, all the ingredients for the BoT simulations can be defined without ambiguity. It is worth noting that, using this procedure, we are able to fix the drive parameter $\eta$ for a single resonance, leading to an intrinsic discrepancy between the original and the mapped system. 

This feature is the most significant difference characterizing the AE scenario and the BoT model, although this limitation could be overcome by extending the linear map of \eref{fkablksdbvlhj} to more complicated nonlinear analogues. This, however, is beyond the scope of the present analysis and, to some extent, beyond the intended aim of any simplified/reduced approach. Adapting the mapping to each single reference would correspond to deal with a ``patch description'' of the one-to-one basic relations. The matching of different patches would imply a very and, under many aspects, redundant
effort. In fact, if the resonances are sufficiently overlapped, the amount of error in describing all the resonances using the single framework of the reference resonance results in a negligible effect. 
However, in the present work, we are focused in showing the capability of our scheme to predict deviations form the QL model, like avalanche phenomena. The adopted paradigm of having a set up based on a reference resonance is valuable to capture, with good accuracy, the specific features of the EP radial transport.

Here, once the drive parameter $\eta$ has been evaluated, additional modes can be included by means of the resonance conditions written in the radial space, \ie
\begin{align}\label{hkjghuii}
\ell_j=\ell^*/(1-s_{rj})\;.	
\end{align}
The dispersion relation \eref{disrel} can then provide, for each mode, its linear drive and frequency. Thus, we can now also map the mode damping as described above, \ie
\begin{align}\label{gammad}
\bar{\gamma}_{dj}/\bar{\omega}_j=\bar{\Gamma}_{dj}/\bar{\Omega}_j\;.
\end{align}
This set of rules completes our description of the map between the radial and velocity spaces, \ie $s\to u$.

\subsection{Map of the Quasi-Linear model}
Let us now implement the mapping procedure to obtain the QL equations directly in the radial coordinate. In particular, we address the QL model in the spirit depicted in \citer{ncentropy} and \citer{ql20nc} for the relaxation of the distribution function and the spectral evolution. Using the linear character of the map \reff{mapvs}, the QL equations can be written for the EP/AE system profile $f_H(\tau,s)$ as
\begin{subequations}\label{QL_s}
\begin{align}
f_H&=F_{H0}-\pi\bar{\mathcal{N}}R\p_s[(1-s)^{-5}\bar{\mathcal{I}}\,]\;,\\
\p_t \bar{\mathcal{I}}&=-\frac{\eta}{R}(1-s)^{2}\,\bar{\mathcal{I}}\,\p_s F_{H0}+
\\
&\qquad\quad+\pi \eta (1-s)^2\bar{\mathcal{N}} \,\bar{\mathcal{I}}\,\p_s^2[(1-s)^{-5}\bar{\mathcal{I}}\,]\;.
\end{align}
\end{subequations}
Here $F_{H0}(s)$ is the initial EP distribution of \eref{fhot} and $R=\int ds F_{H0}$. We have also used the following normalization for the spectrum $\bar{\mathcal{I}}(\tau,s)=(\ell^*)^{5}|\bar{\phi}|^2/\eta$, where $\bar{\phi}(\tau,s)$ has to be intended as the continuous spectrum derived from the discrete one, specified by means of the relation \reff{hkjghuii}. Moreover, we have defined the spectral density as $\bar{\mathcal{N}}=M/\Delta\ell$ (where $\Delta\ell=\textrm{Max}[\ell]-\textrm{Min}[\ell]$ is the spectral width).

\section{Numerical results}\label{sec_sims}
In this Section, we apply the mapping procedure defined above to the specific case of the ITER 15MA beseline scenario of \citer{spb16} (see also \citers{ML12,ML13,Ph15}). By means of the proper parameter setup, we run simulations of \erefs{mainsys1} and then we map back the results into the normalized radial dimension. Moreover, using \erefs{QL_s}, we are able to compare the non-linear and QL evolutions of the distribution function. 

\subsection{Parameter setup and linear analysis}
\begin{figure}[ht!]
\centering
\includegraphics[width=.4\textwidth,clip]{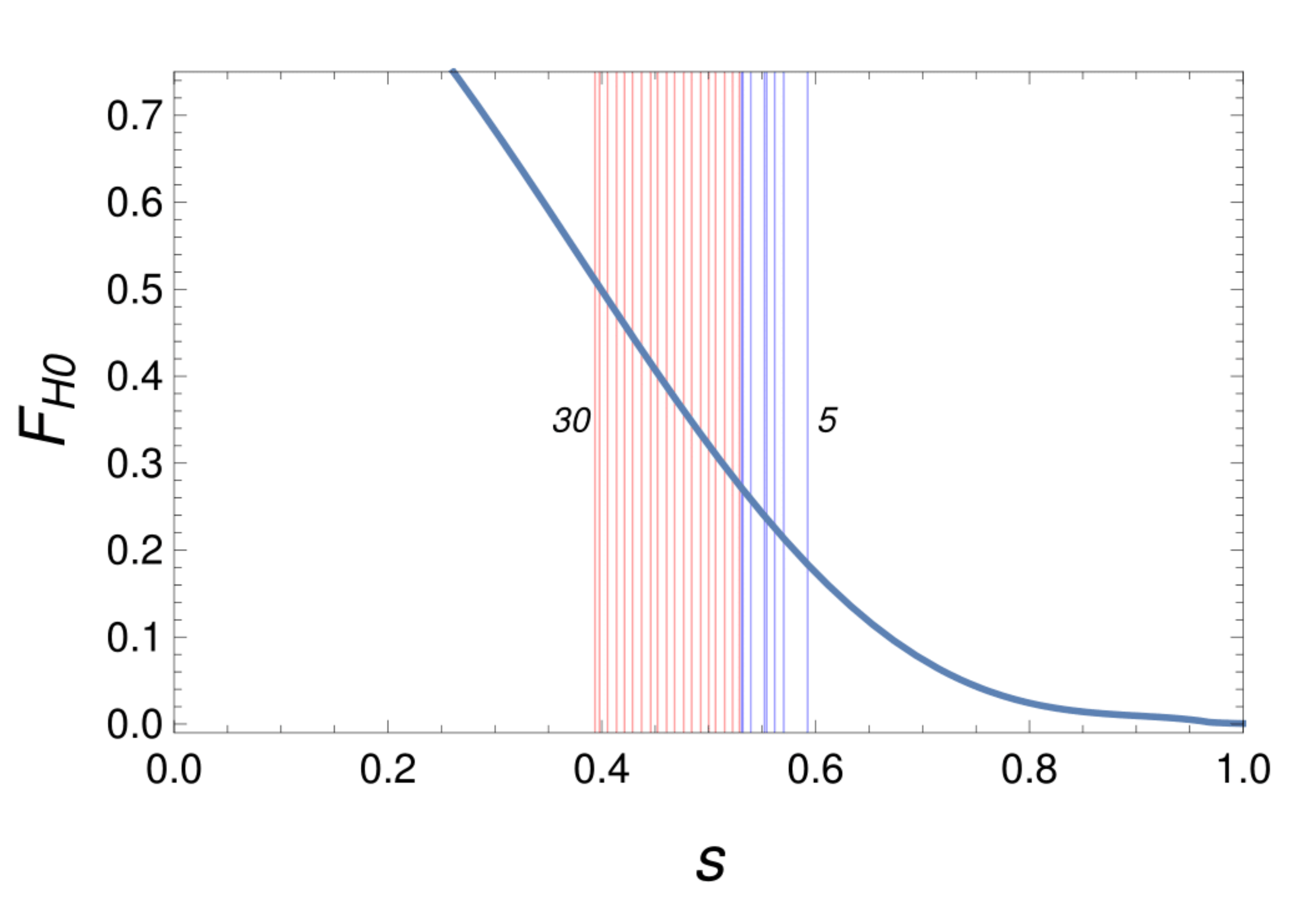}\\
\includegraphics[width=.4\textwidth,clip]{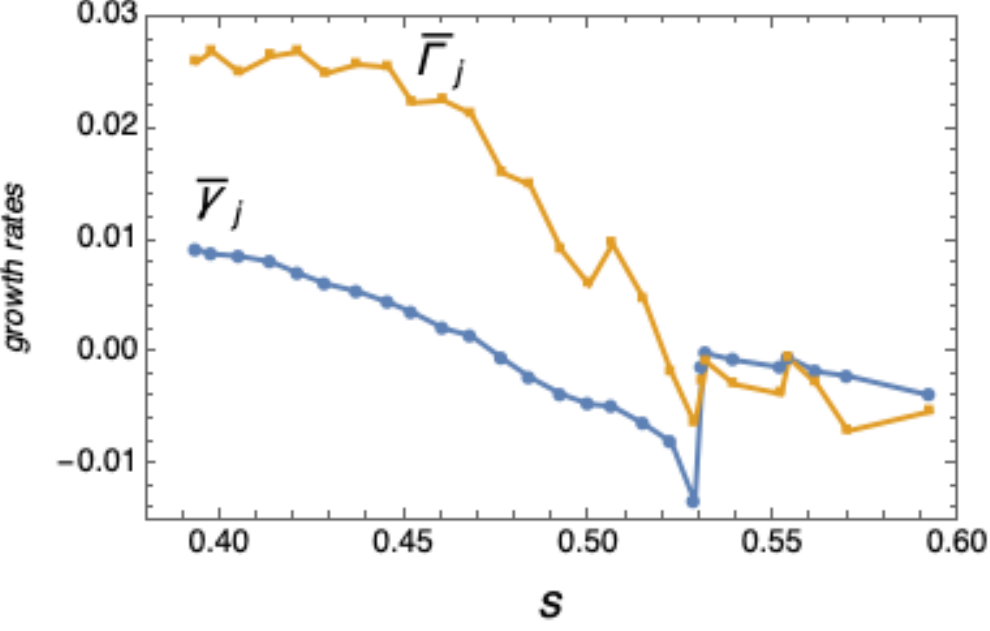}
\caption{(Color online) Upper panel - Initial EP profile $F_{H0}$ (arbitrary units) as a function of the normalized radius $s$. Colored lines represent the resonance positions of the analyzed modes: high-$n$ branch in red, low-$n$ branch in blue. Integer numbers indicate the toroidal mode numbers $n$ for the extrema. \;\;Lower panel - Plot of the growth rates $\bar{\gamma}_j$ and $\bar{\Gamma}_j$ (data from \citer{spb16}) as indicated in the figure, as a function of the radial position of the corresponding mode. $\bar{\gamma}_j$ are evaluated from \eref{disrel} using \eref{hkjghuii} and \eref{gammad}.
\label{figsetup}}
\end{figure}

The scenario analyzed in \citer{spb16} is specified for the initial EP profile described by \eref{fhot} with $B_1\simeq0.5$, $B_2\simeq-1.2$ and $B_3\simeq3.2$, and it includes the least damped 27 TAE modes defined from the linear analysis of \citer{Ph15} (see the upper panel of \figref{figsetup}, where we plot $F_{H0}(s)$ and we indicate with colored lines the mode resonance positions).  In particular (see \citer{spb16} for a detailed discussion on this choice), $n\in[12,30]$ for the main branch (red) and $n\in[5,12]$ for the low branch (blue) are treated (we discuss the role of the poloidal harmonics in the following Sec.\ref{sec_har}). As reference resonance, $n=21$ of the main branch has been chosen since it is characterized by the average value of radial position and essentially also of the growth rate of the branch (as far as $\bar{\Omega}_F(s)$ is linear enough, the choice of the reference resonance does not significantly affects the obtained profiles). 

The parameter $\alpha$ can be set in order to obtain optimized EP transport in the $s$ direction, with respect to the 3D relaxation. In \citer{spb16}, it was pointed out how the EP fully self consistent redistribution taken at times just after saturation actually matches the QL predicted profiles. Only after this phase of the evolution an avalanche process towards the edge occurs. In this respect and for the purpose of the present work, we now set the scaling parameter of \eref{scaling} by performing a scan on $\alpha$ in order to match the relaxed QL flattening width of the two schemes by straightforward integration of \erefs{QL_s} (it is clear that that this procedure can actually be implemented by a direct scan on the $\eta$ parameter, but we stress how the relevant physical insight is contained in \eref{scaling} since it provides us with an additional information roughly concerning the fraction of the fully resonant particle of the realistic 3D scheme). Actually, the QL evolution presented \citer{spb16} outlines a flattening width between $s\simeq0.34$ and $s\simeq0.6$ which is well reproduced by $\alpha=0.4$ corresponding to $\eta\simeq0.007$. For the sake of completeness, we also mention that, for the simulations, we have set $\ell^*=400$. We remark that, for this stage of the mapping procedure, the definition of $\alpha$ remains a critical point that we discuss in details in Sec.\ref{alpha_sec}, where we also outline a strategy to define the mapping without drive scaling.

Integrating the dispersion relation \reff{disrel} using \eref{hkjghuii} and setting the damping profile via \eref{gammad}, we obtain the growth rates $\bar{\gamma}_j=\bar{\gamma}_{Dj}-\bar{\gamma}_{dj}$ for each mode as plotted (vs the resonant radius) in the lower panel of \figref{figsetup}, where we also plot the corresponding values $\bar{\Gamma}_j=\bar{\Gamma}_{Dj}-\bar{\Gamma}_{dj}$ for the TAEs. As it emerges from the figure, the profiles of the growth rates appear reliable (remembering the scaling $\alpha=0.4$) for the main branch. Instead a discrepancy appears for the low-$n$ branch, which in our case has a marked discontinuity with respect to the high-$n$ region. Actually, in such a domain the frequency of the TAEs changes significantly from one branch to the other. This feature, combined with an opposite discontinuity in the corresponding values of the damping rates, generates a rather smooth profile for $\bar{\Gamma}_j$ (see \citers{Ph15,spb16} for the detailed linear analysis). This effect is not reproduced in the 1D model, for which the mode frequencies remain essentially $\omp$.
\begin{figure}[ht!]
\centering
\includegraphics[width=.5\textwidth,clip]{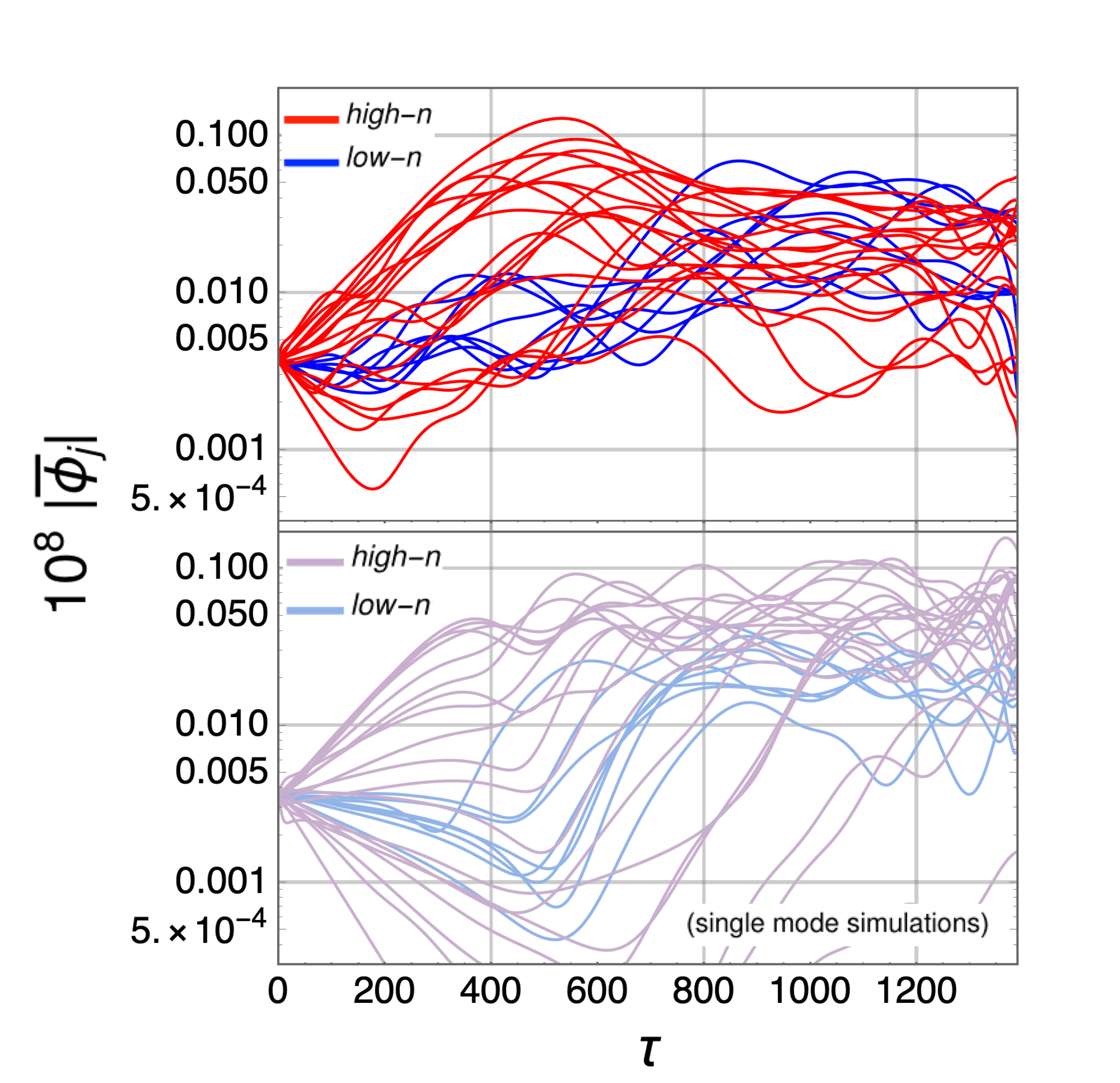}\vspace{-0cm}
\caption{(Color online) Self-consistent evolution of the 27 modes (bright colors - upper subplot) and the correspondent single mode behaviors (opaque colors - lower subplot). Blue represents the low-$n$ branch, while red the high-$n$ branch, as indicated in the figure.
\label{figmode}}
\end{figure}

\subsection{Non-linear simulations}
Let us now discuss the non-linear results obtained from simulations of \erefs{mainsys1} and plotted in the mapped $s$ domain. In \figref{figmode}, we depict the mode evolution by comparing the multi mode simulation in bright colors - upper subplot (one simulation in which all the 27 modes are self consistently evolved together) with respect to the single mode evolution in opaque colors - lower subplot (27 distinct simulations evolving one mode individually). The low-$n$ branch (blue) results to be more efficiently exited in the simultaneous presence of the modes (compared to the single-mode cases) especially during the early linear phase around $\tau\simeq300$. This feature is due to an expected avalanche transport of particles in the correspondent resonance radial portion. Such a result is consistent with \citer{spb16}, where the high-$n$ branch grows faster and the low-$n$ branch is dominant also in the nonlinear phase: a domino-like behavior emerges. Moreover, we are able to reproduce a peculiar feature of the spectrum: the saturation level of some modes is larger than that of the corresponding single mode evolution. This shows the strong interplay effect of the multi-mode regime. The capability of our mapping procedure to predict and reproduce such specific features (nonetheless they are less evident in our case due the adopted scaling for the drive) constitutes a significant validation of our methodology.


%
%

\begin{figure}[ht!]
\centering
\includegraphics[width=.4\textwidth,clip]{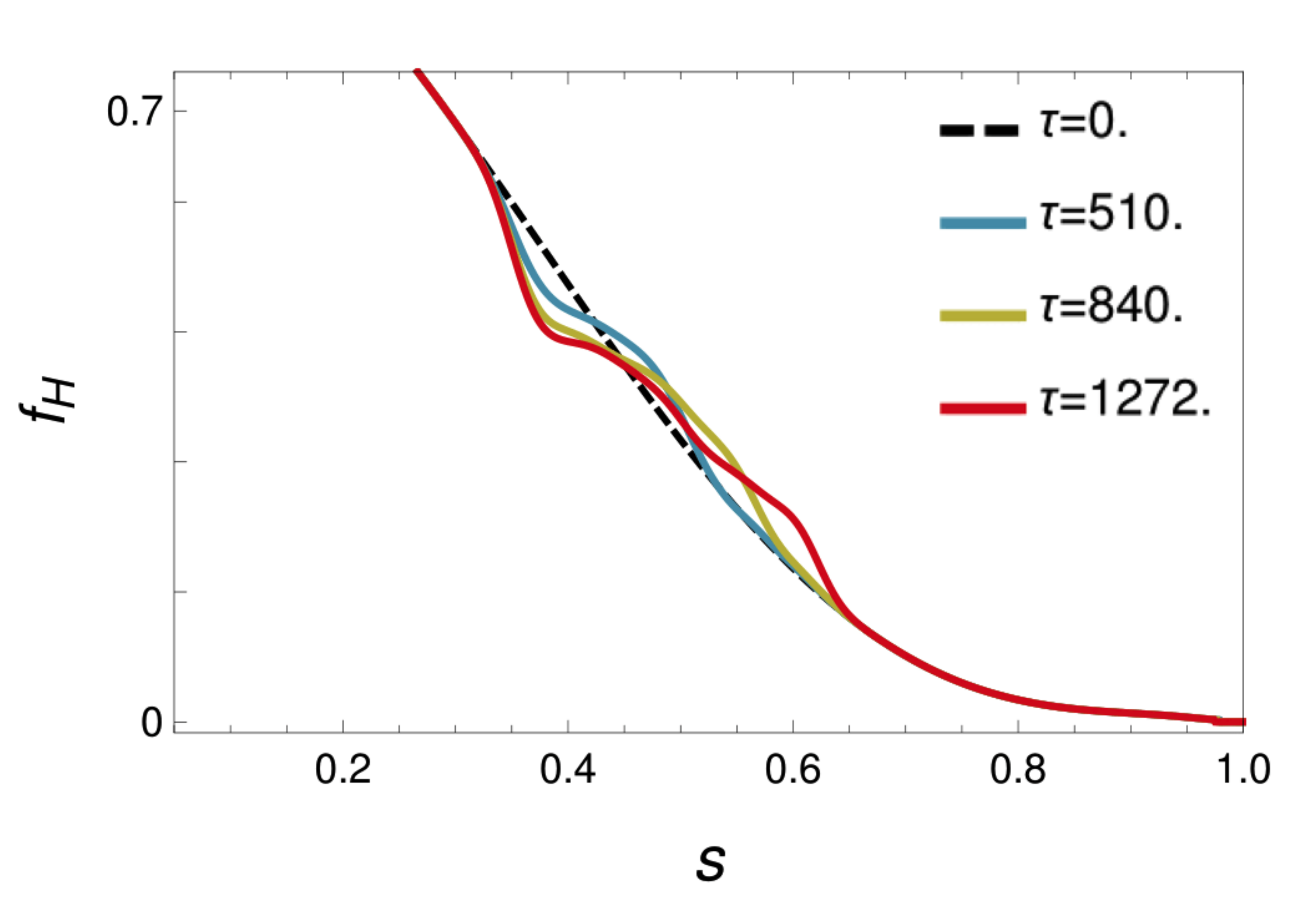}\\
\includegraphics[width=.395\textwidth,clip]{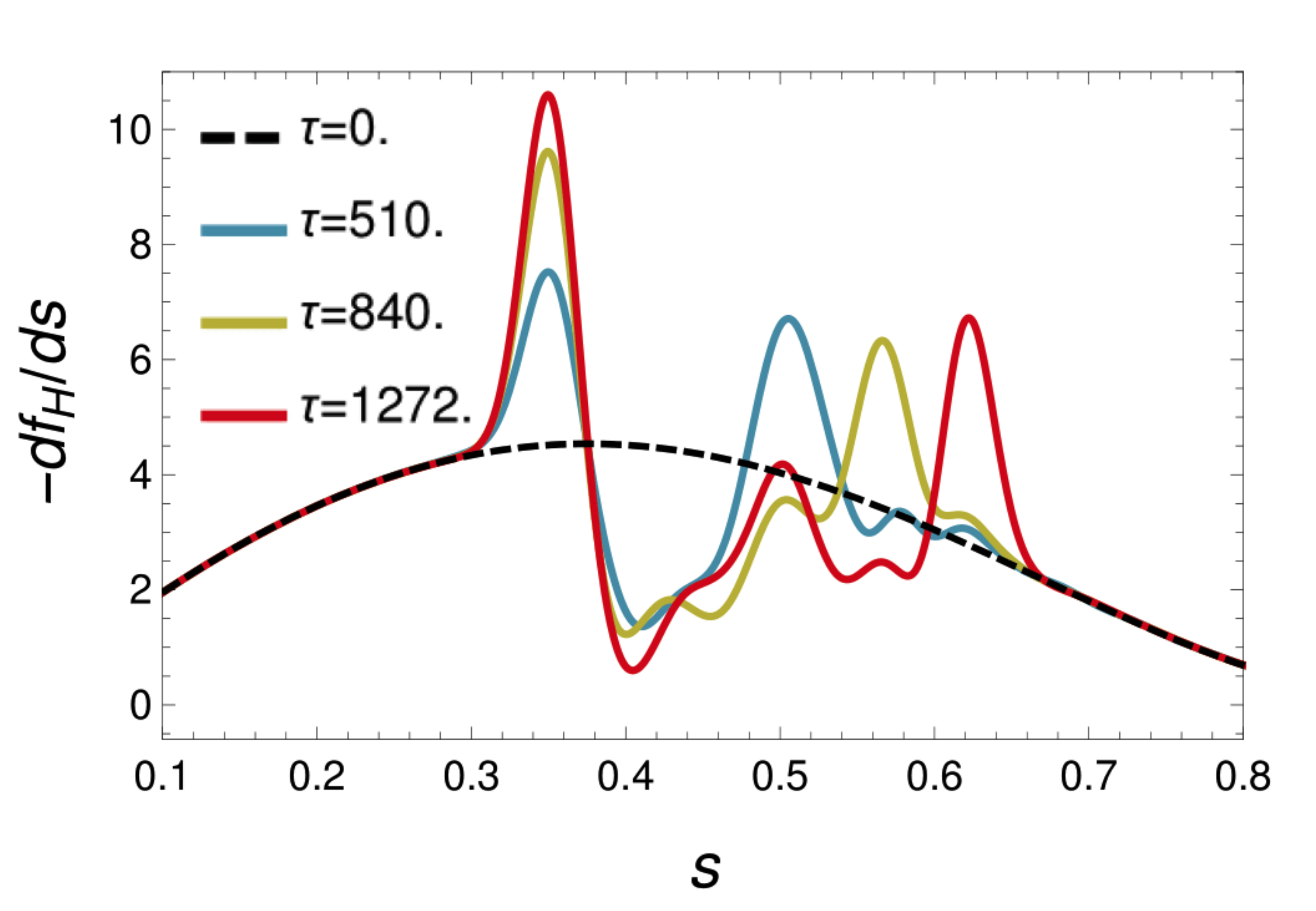}
\caption{(Color online) Upper panel - Evolution of EP profile (arbitrary units) as function of the normalized radius $s$. Colored lines represent different times, as indicated in the plot, while the dashed black line is the initial distribution $F_{H0}$.
\;\;Lower panel - Plot of the opposite radial gradient of the profile, \ie $-df_H/ds$, using the same colors.
\label{figfh}}
\end{figure}

The spectral evolution affects the EP redistribution and the evolution of the distribution function $f_H(s)$ (upper panel) ad its radial gradient (lower panel) are plotted in \figref{figfh}, where transport toward large radius clearly emerges. In particular, while a peak in the gradient of EP profile in well localized around $s\simeq0.35$ (characterizing the excitation of the high branch), a second peak, related to the low branch instability, is shifting in time toward $s\simeq0.65$ outlining the convective-like transport toward the plasma edge. When considering the radial gradient of the profiles, we stress how the growing of the first peak reflects the evolving form of the distribution function which is gradually reaching a flat configuration. This is due the emptying of the phase space related to the corresponding mode enhancement and, thus, to energy conservation.
\begin{figure}[ht!]
\centering
\includegraphics[width=.4\textwidth,clip]{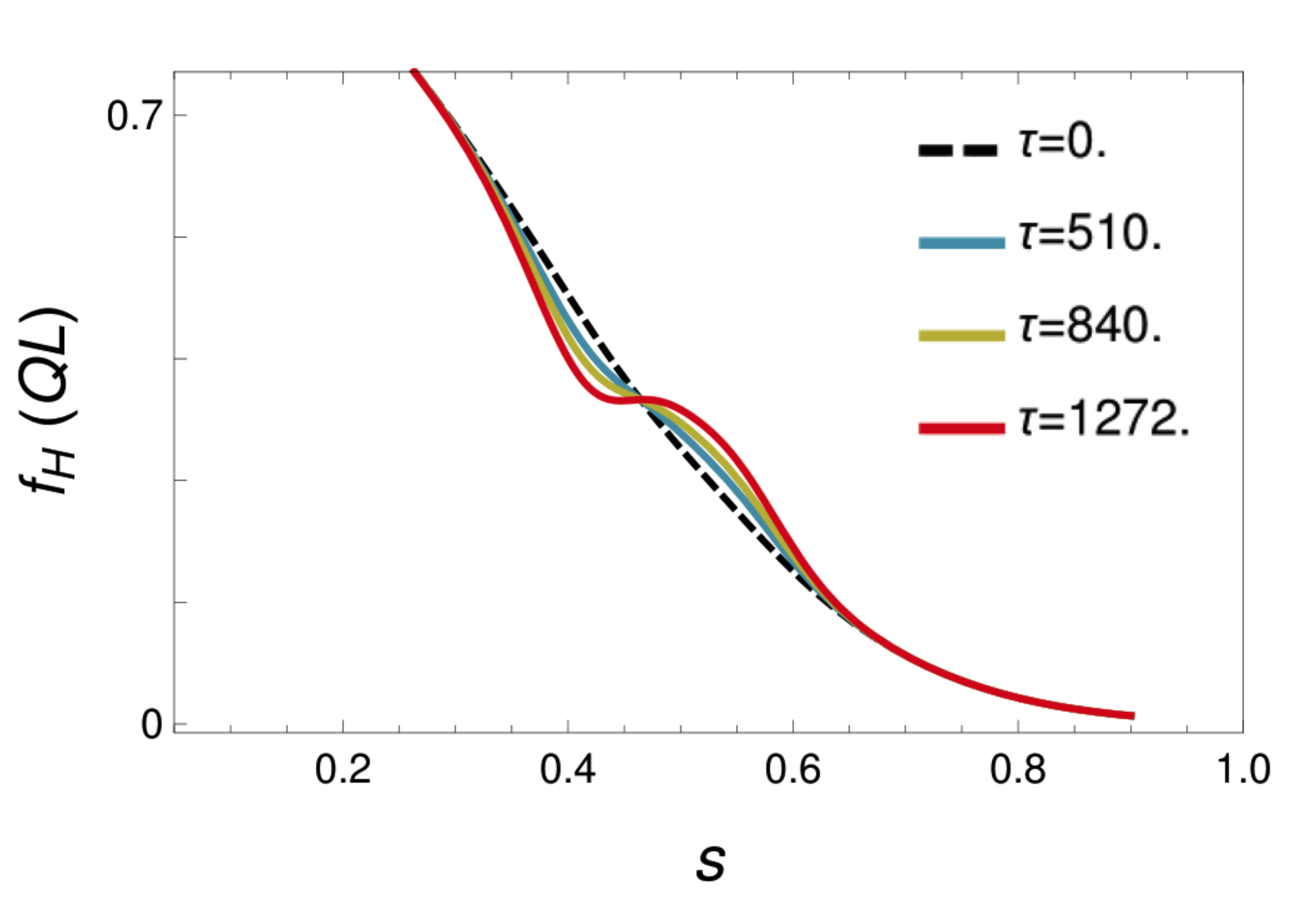}\\
\includegraphics[width=.395\textwidth,clip]{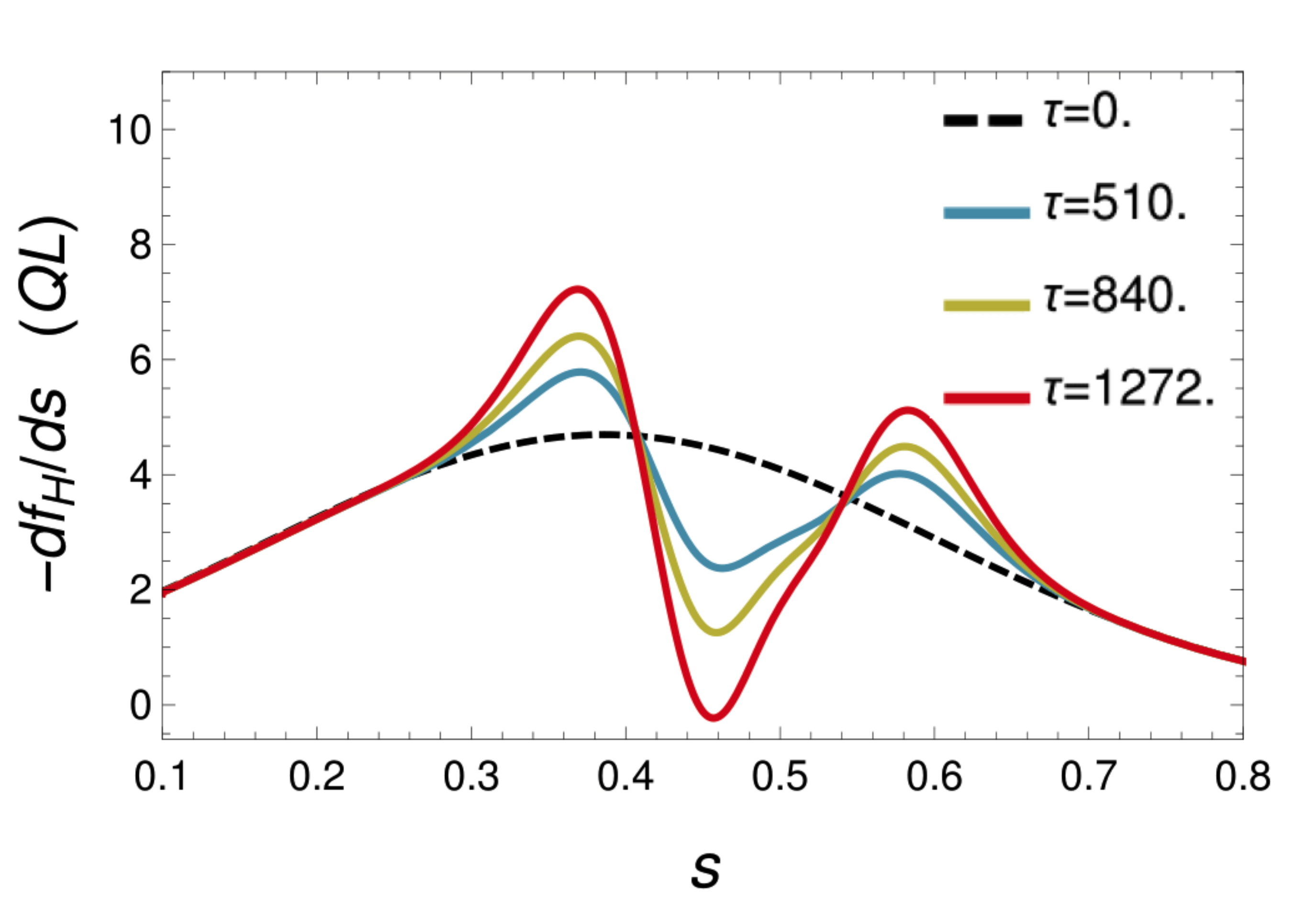}
\caption{(Color online) Upper panel - Radial EP redistribution as derived from QL theory, by integrating \erefs{QL_s}. Colored lines represent different times, accordingly to \figref{figfh}.\;\;
Lower panel - Evolution of the radial gradient of the distribution function evaluated from QL model, using the same color scheme.
\label{figfhql}}
\end{figure}

\subsection{Comparison with respect to QL transport}
Let us discuss the differences of the previous analysis versus the QL evolution in close analogy to \citer{spb16}. \erefs{QL_s} are numerically integrated using the same initial EP distribution $F_{H0}$ of the simulations. As initial condition for $\bar{\mathcal{I}}(\tau,s)$, we adopt a Gaussian profile which model the discrete mode spectrum, \ie which is localized in the resonance region represented in the upper panel of \figref{figsetup} and with an amplitude which mimics the linear phase of the mode evolution. Results are plotted in \figref{figfhql}, outlining a clear diffusive flattening which, as already discussed, by construction matches the radial spread of  \citer{spb16}. The discrepancy with respect to the self-consistent evolution is now  evident. In particular, the avalanche phenomenon and the corresponding radial transport is absent. This further elucidates the importance of the convective character of the dynamics pointed out by self-consistent simulations, which is not reproducible with a pure diffusive model as the QL theory. The two peaks of the distribution radial gradient are now well radially localized during the system evolution and the outer redistribution of the EP is suppressed.

\subsection{Test particle transport: diffusion vs convection}\label{sec_conv}
In this subsection, we focus on the transport character during the self-consistent relaxation process. In particular, we analyze the test particle evolution under the influence of extracted self-consistent potential fields from the simulation of \figref{figfh}. The tracers can be initialized in radius (velocity) to represent different portions of the distribution function. In order to define the transport features, for a given initial condition around a reference value $s_0$, we analyze the mean square path of the test particles as a function of time:
\begin{align}\label{jkjhfbfy}
\langle\delta s^2\rangle=\langle[s(\tau)-\langle s(\tau)\rangle]^2\rangle\;,
\end{align}
here, the average $\langle...\rangle$ is taken over the tracers. If $\langle\delta s^2\rangle$ outlines a linear dependence in $\tau$, this clearly indicates diffusive transport, while convection is characterized by $\langle\delta s^2\rangle\propto \tau^2$. We stress how this study is, in general, non trivial: $\langle\delta s^2\rangle$ can be influenced by complicated nonlinear time evolution when tracers are significantly displaced in the phase space (a detailed analysis of this issue can be found in \citer{VK12}, see also \citer{ncentropy} for a related study), thus generating different and peculiar profiles of the mean square path.
\begin{figure}[ht!]
\centering
\includegraphics[width=.4\textwidth,clip]{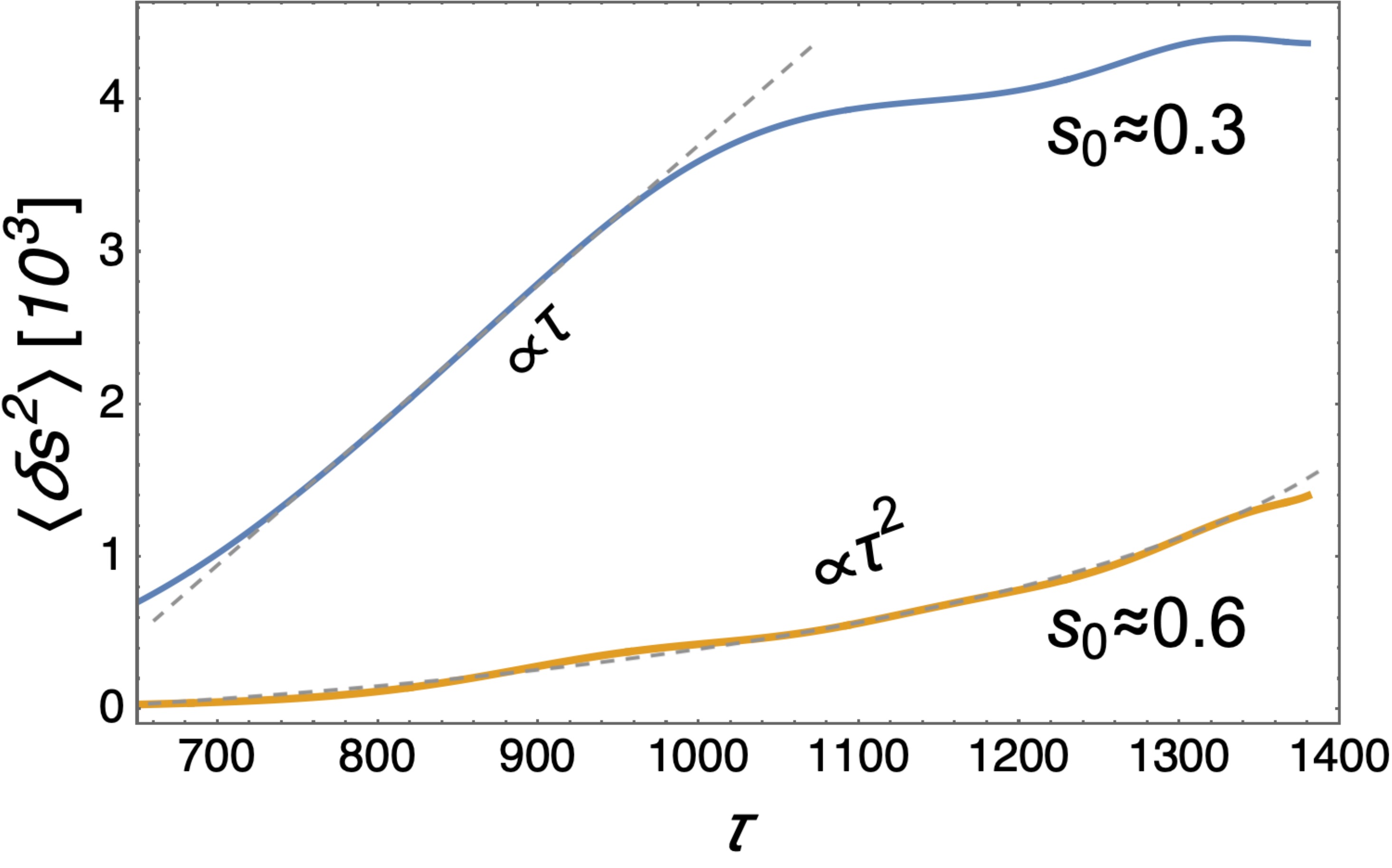}
\caption{(Color online) Plot of $\langle\delta s^2\rangle$ from \eref{jkjhfbfy} as a function of time for test-particles initialized at $s_0\simeq0.3$ (blue) and around $s_0=0.6$ (yellow). Dashed gray lines are guide for eyes underlining the different transport schemes. 
\label{figmmsp}}
\end{figure}

In \figref{figmmsp}, we plot $\langle\delta s^2\rangle$ as a function of time for tracers initialized around $s_0=0.3$, characterizing the relaxation of the first peak of \figref{figfh}, and around $s_0=0.6$, thus related to the second peak (this value indeed corresponds to the peak localization of the QL model). Focusing on time scales after saturation (\ie after the dynamics drive dominance by coherent rotating clumps in the phase space), the diffusive behavior of the first peak is clearly evident during the post saturation phase. Therefore, the shape and the width of the relaxed profile is predicted by the QL evolution with a reasonable degree of precision. The late dynamics does not outline pure diffusive feature due to the statistical weight of significantly displaced tracers (see also \citers{VK12,ncentropy}). The evolution of the second peak (tracers initialized at $s_0\simeq0.6$) is instead characterized by a $\tau^2$ dependence, during the whole considered time interval, underlining the convective character of the transport. This analysis actually confirms the evidences outlined in the previous sub-section. In fact, nonetheless the convection rate results lower than the diffusive counterpart, the second peak transport can not be modeled by the QL treatment, as already discussed, due to the dominance of avalanche transport related to the low-$n$ branch.

\subsection{The role of harmonics}\label{sec_har}

Concerning the self-consistent evolution of \figref{figfh}, it is important to stress that, since we are addressing the dynamics of a fixed set of modes, the transport process is bounded at $s\simeq0.65$ due to the absence of further modes to be excited via the avalanche mechanism. In this respect, in \citer{spb16} an outer redistribution of the EP profile has been shown to be triggered toward $s\simeq0.85$. Such a discrepancy with respect to our analysis is due to the importance of the poloidal harmonics spectrum. In fact, in \citer{spb16}, all the harmonics with a peak greater the 25\% of the mode maximum peak were considered, thus the low branch has been simulated with 12 poloidal harmonics, while the high branch is characterized by 2 harmonics only (we refer to the paper for other details). These additional modes involve a wider resonant region \cite{Ph15} and this generates a further transport to the edge. We now discuss a model to include the effects of the harmonic spectrum in the non-linear simulations.
\begin{figure}[ht!]
\centering
\includegraphics[width=.4\textwidth,clip]{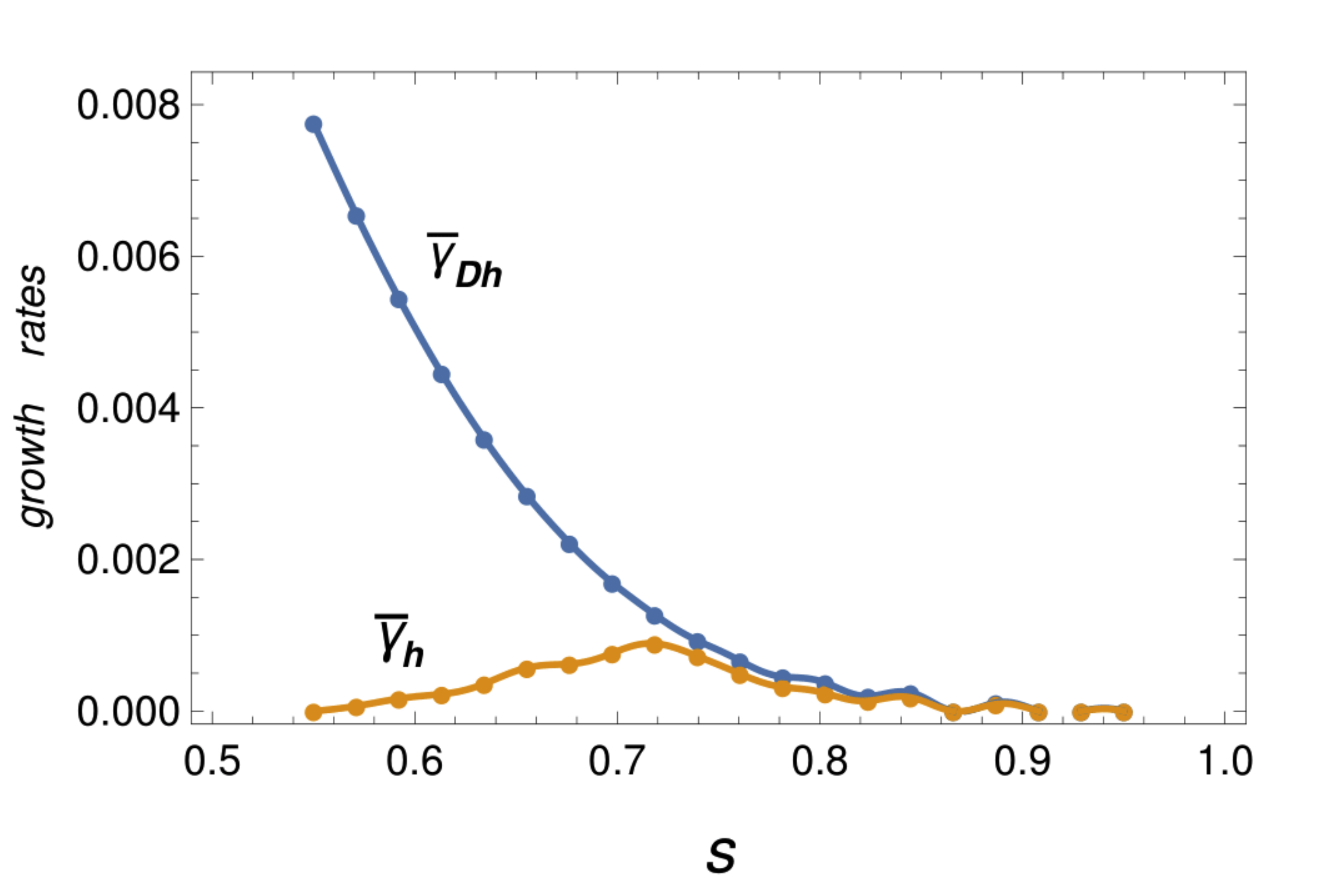}
\caption{(Color online) Plot of $\bar{\gamma}_{Dh}$ (blue) and $\bar{\gamma}_{h}$ (yellow) for the additional 20 modes, as function of the resonance radius.
\label{figharms}}
\end{figure}
\begin{figure}[ht!]
\centering
\includegraphics[width=.4\textwidth,clip]{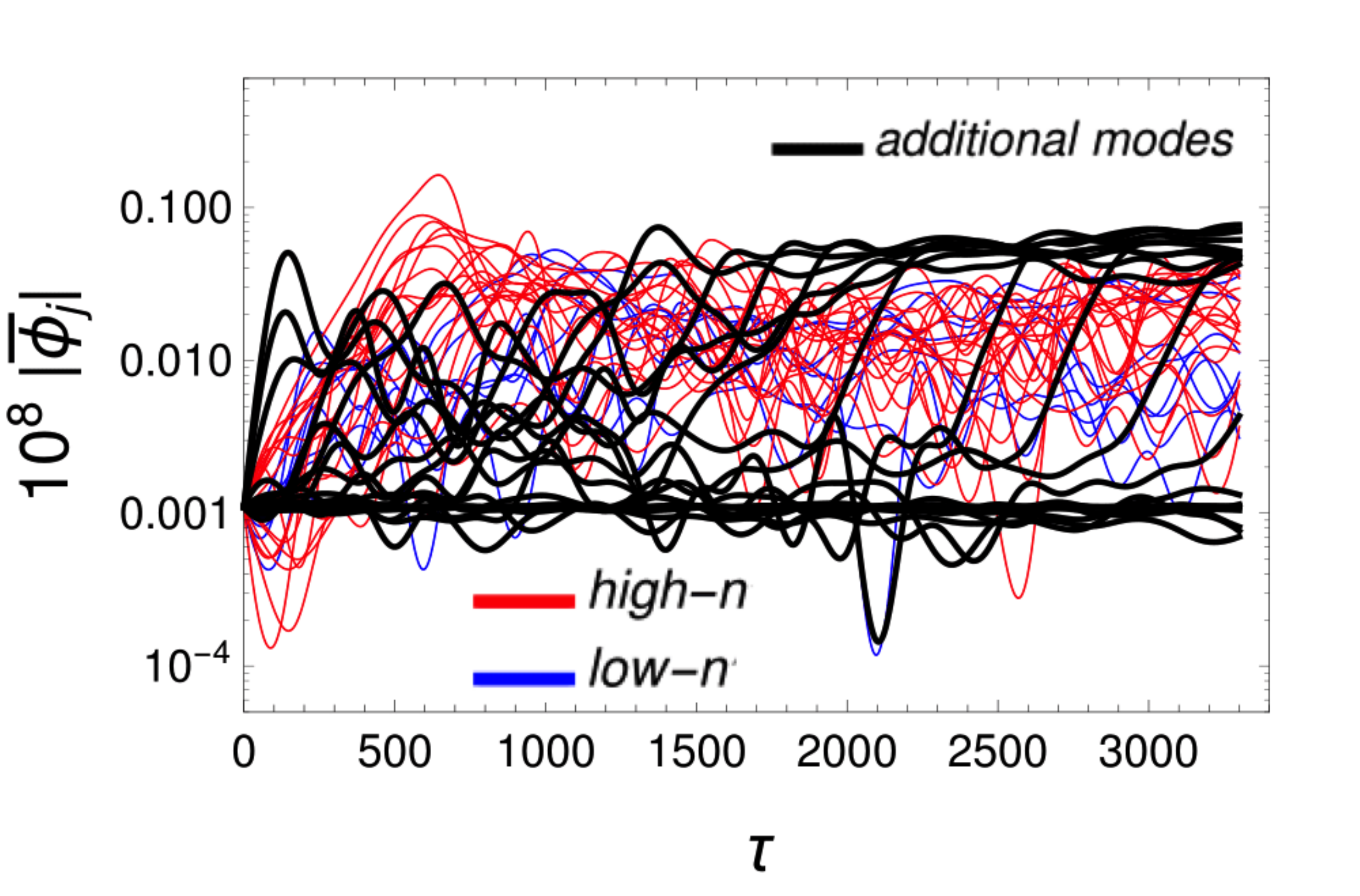}\\
\includegraphics[width=.37\textwidth,clip]{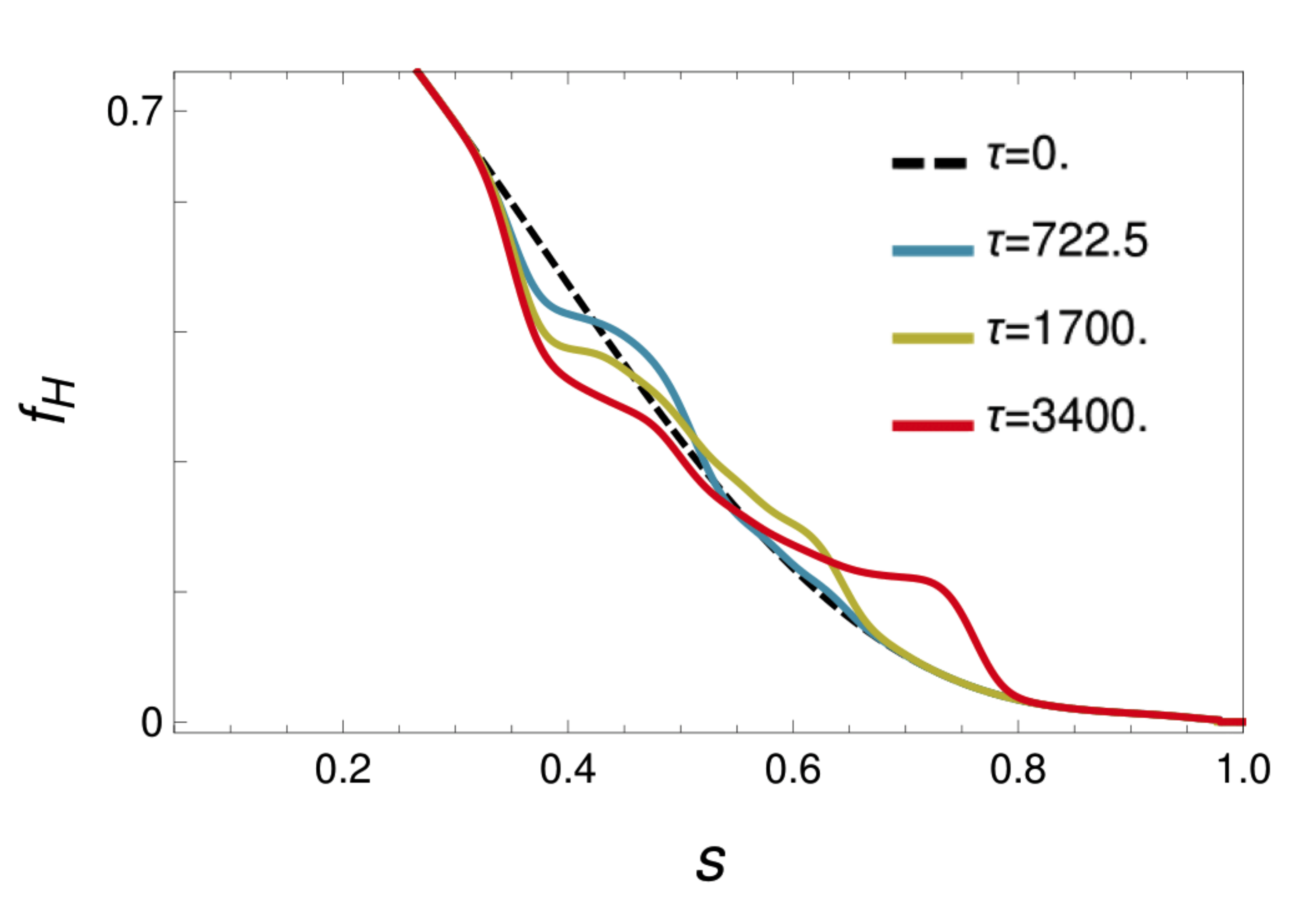}
\caption{(Color online) Upper panel - Self-consistent mode evolution: blue (red) denotes low (high) -$n_{AE}$ branch, while the added 20 modes defined in \figref{figharms} are indicated in black. Lower panel - Evolution of EP distribution function versus $s$. Colored lines indicate distinct times (dashed black line is the initial distribution $F_{H0}$).
\label{figharmsmodfh}}
\end{figure}

As already mentioned, in the linear analysis of \citer{Ph15}, it has been pointed out how the poloidal harmonics fill a wide resonance region. In this respect, we try to reproduce this feature by adding other $20$ modes having wave numbers resonating with $0.55\leqslant s \leqslant 0.95$. Given the linear parameter setup defined above, each additional mode has a linear drive $\bar{\gamma}_{Dh}$ (with $h=1,\,...,\,20$) defined by the dispersion relation \eref{disrel}. In order to reproduce the proper morphology of the harmonic spectrum, we consider specific damping rates $\bar{\gamma}_{dh}$ yielding a Gaussian distribution for $\bar{\gamma}_{h}=\bar{\gamma}_{Dh}-\bar{\gamma}_{dh}$. Imposing the constraint of having $\bar{\gamma}_{dh}\geqslant0$, the resulting shape of the growth rates is depicted in \figref{figharms} together with $\bar{\gamma}_{Dh}$.

Simulation results are given in \figref{figharmsmodfh}. In the upper panel, we depict the mode evolution, where the spectral transfer clearly emerges. The additional modes are, in fact, progressively excited resulting in an increased avalanche transport toward large radial positions. This is clearly represented in the lower panel, where an outer redistribution is triggered around $s\simeq0.75$. Comparing these results with the analysis in \citer{spb16} (outer redistribution at $s\simeq0.85$), it is evident how the domino transport is less efficient in our case, due to the fixed character of the spectrum which can not precisely mimic the harmonics effects near the edge. Moreover, this feature intrinsically gives rise to small differences also in the shape of the distribution function. Despite the agreement of our simulations with the one presented in \citer{spb16} is on a qualitative level, the main merit of the present study is to shed light on the real mechanism responsible for the non-pure diffusive transport, predicted by the QL model. Actually, the presence in the spectrum of neighboring resonances (like the poloidal spectrum discussed above) induces EP avalanches that enhance the transport toward the edge.

\section{Reduction from 3D to 1D}\label{alpha_sec}
As already mentioned, the distribution function evolution described in the previous Section is compared to the global radial redistribution of the realistic 3D case. In this sense a scale factor $\alpha$ is introduced to take into account the reduced dimensionality of the BoT model, see \eref{scaling}. This is due to the fact that the realistic radial profiles are integrated over the whole EP velocity space distribution function, including resonant as well as non-resonant particle. Meanwhile, the fraction of resonant particles reflects the equilibrium geometry as well as the number of degrees of freedom of single particles interacting with the fluctuation spectrum. In this Section, we elucidate the physical nature of the considered drive scaling by showing how the redistribution of a chosen set of resonant EP tracers which maximize the power exchange can be properly described by the mapping procedure without scaling the drives, \ie by setting $\alpha=1$.

In the case of a single resonance, for the realistic 3D case, two constants of motion can be defined \cite{ZCrmp,ZC15njp,BW14pop,white12}: $K=E-\Omega P_\varphi/n$ (where $E$ is the particle energy and $P_\varphi$ the toroidal angular momentum) and the magnetic moment $\mu$. Fixing the constants of motions corresponds to cut the phase space into infinitesimal slices that do not mix together also nonlinearly. In this sense, the dynamics of a single slice results in a 1D system that can be properly described by the BoT reduced model. If the slice is chosen in order to be made of resonant EPs that maximize the power exchange, the mapping procedure has to be be applied without effective drive scaling. 

\begin{figure}[htbp]
\centerline{\includegraphics[width=6cm]{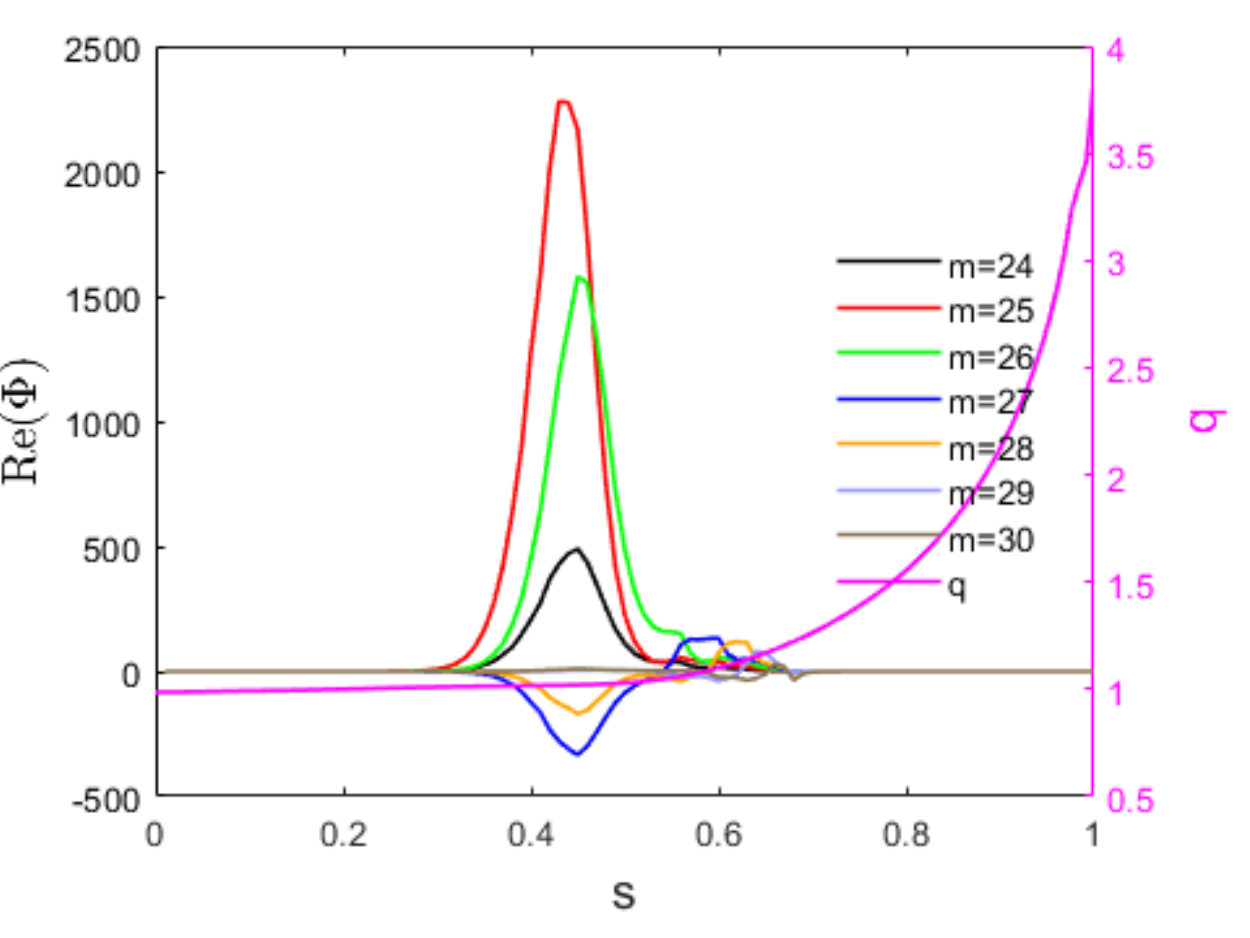}}
\caption{(Color online) The left axis corresponds to the radial mode structure of the $n=25$ TAE; the right axis corresponds to the safety factor $q$.}
\label{fig:qprofile}
\end{figure}

In the following, we describe the procedure used to properly identify such a most resonant slice for a test case, namely $n=25$. As shown in \figref{fig:qprofile}, the dominant poloidal harmonics of the considered TAE are $m=25,\,26$ and the mode peaks at $s_r=0.43$. 
The mode width is about 0.2 while the mode frequency is $83.85$\,kHz. The damping rate calculated by LIGKA \cite{lauber2007ligka} is $\bar{\Gamma}_d/\bar{\Omega}=0.923\%$.

The self-consistent simulation of EP driven AE is done by HAGIS code \cite{pinches1998hagis}. This code models the EP-waves nonlinearity via the interaction Lagrangian while other nonlinearities are ignored. The mode structure is fixed and the damping rate is an input parameter. We use the already discussed ITER 15MA scenario equilibrium with realistic alpha's density profiles \cite{Polevoi2002}. The pitch angle distribution of the alpha particles is isotropic and we use a slowing down energy distribution 
\begin{align}
f_0(E)=\frac{1}{E^{3/2}+E_c^{3/2}} \textrm{Erfc} \left(\frac{E-E_0}{\Delta E}\right)\;,
\end{align}
with a birth energy of $E_0=3.5$\, MeV, $E_c=816$\,keV and $\Delta E=491$\,keV. The volume averaged alpha's beta is $0.17\%$.

\begin{figure}[htbp]\centering
\includegraphics[width=0.4\textwidth]{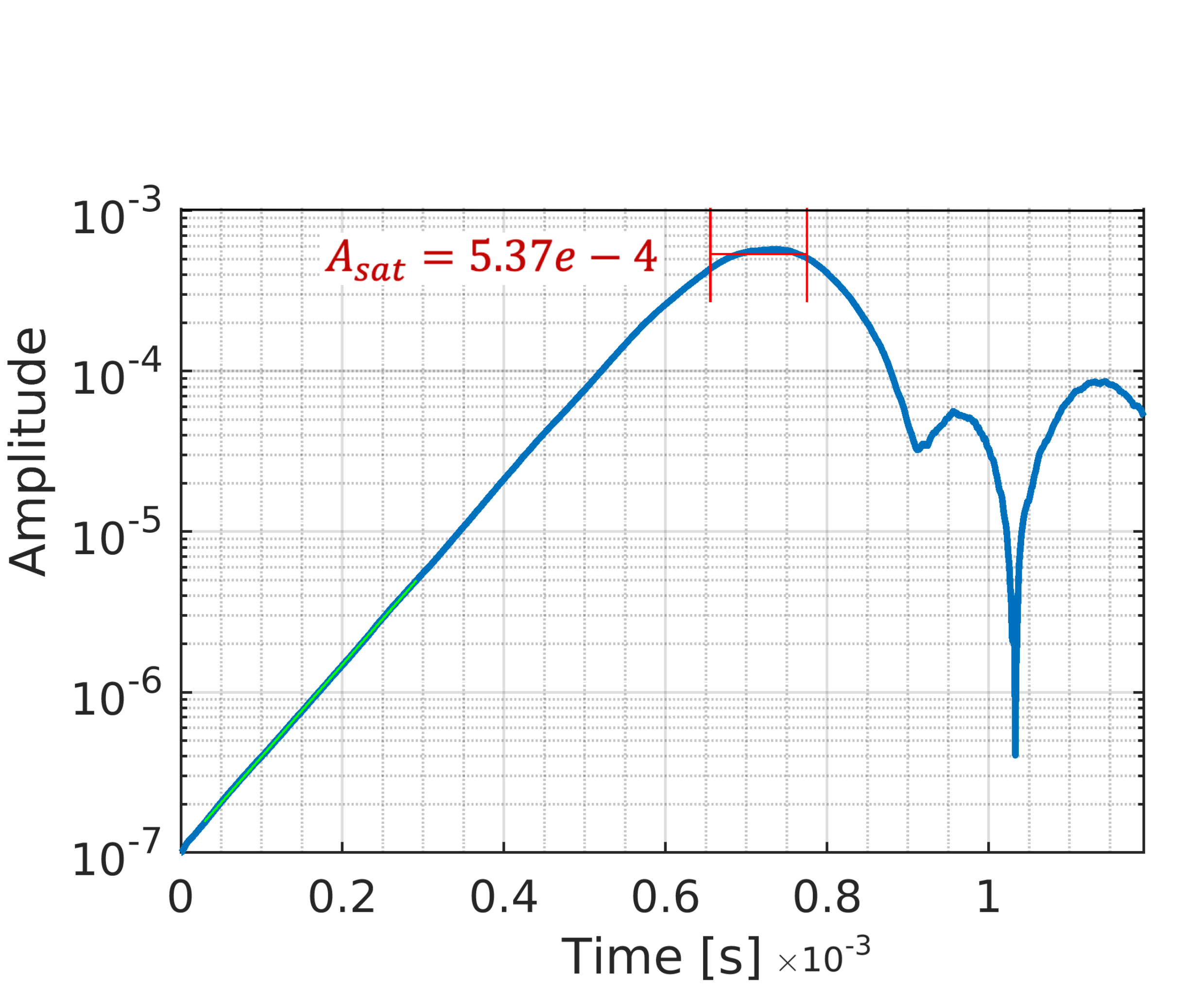}
\caption{(Color online) Mode amplitude $\delta B/B_{mag}$ variation with time using the HAGIS code.}
\label{fig:sat}
\end{figure}

\begin{figure}[htbp]\centering
\includegraphics[width=0.5\textwidth]{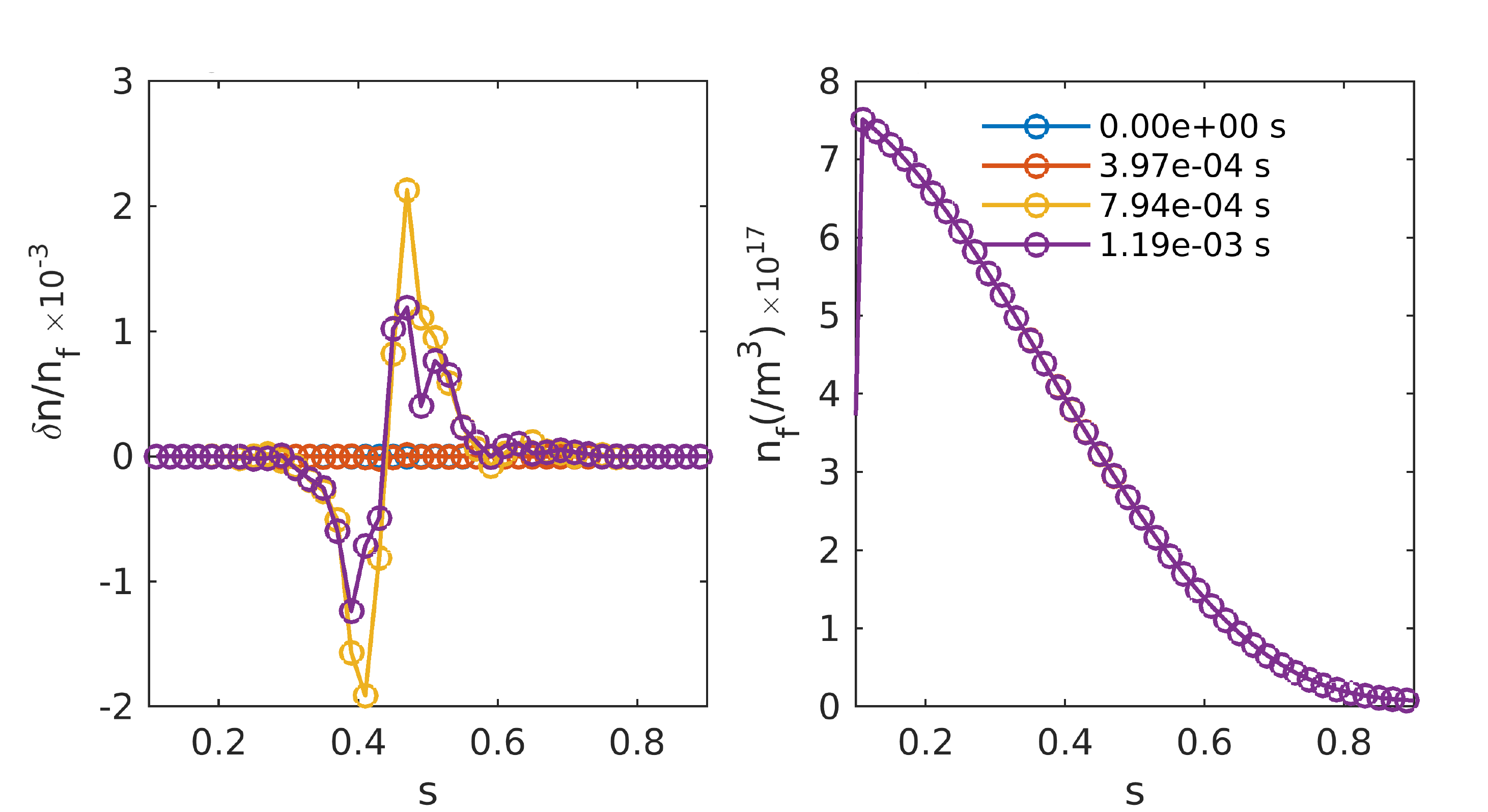}
\caption{(Color online) Time evolution of the EP distribution variation $\delta n$ (left-hand panel) and of the EP density (right-hand panel) using the HAGIS code. The negligible magnitude of $\delta n/n_f$ reflects in a not appreciable modification of the distribution, which remains almost constant during the evolution (all the different times are overlapped).}
\label{fig:del_f}
\end{figure}

The time evolution of the perturbed mode is shown in \figref{fig:sat}. The mode is saturated at $t\simeq 74$ms. The linear growth rate is $\bar{\Gamma}/\bar{\Omega}=2.50\%$ and the saturation level is $A_{sat}=5.37\times 10^{-4}$. This result matches a previous study as shown 
in \citer{spb16}. The EP density variation $\delta n$ at saturation occurs at $s\in[0.3,0.6]$ as shown in \figref{fig:del_f}, but the corresponding magnitude of $\delta n/n_f$ result negligible and the flattening of EP total density ($n_f$) is thus not appreciable.

\begin{figure}[htbp]\centering
\includegraphics[width=0.48\textwidth]{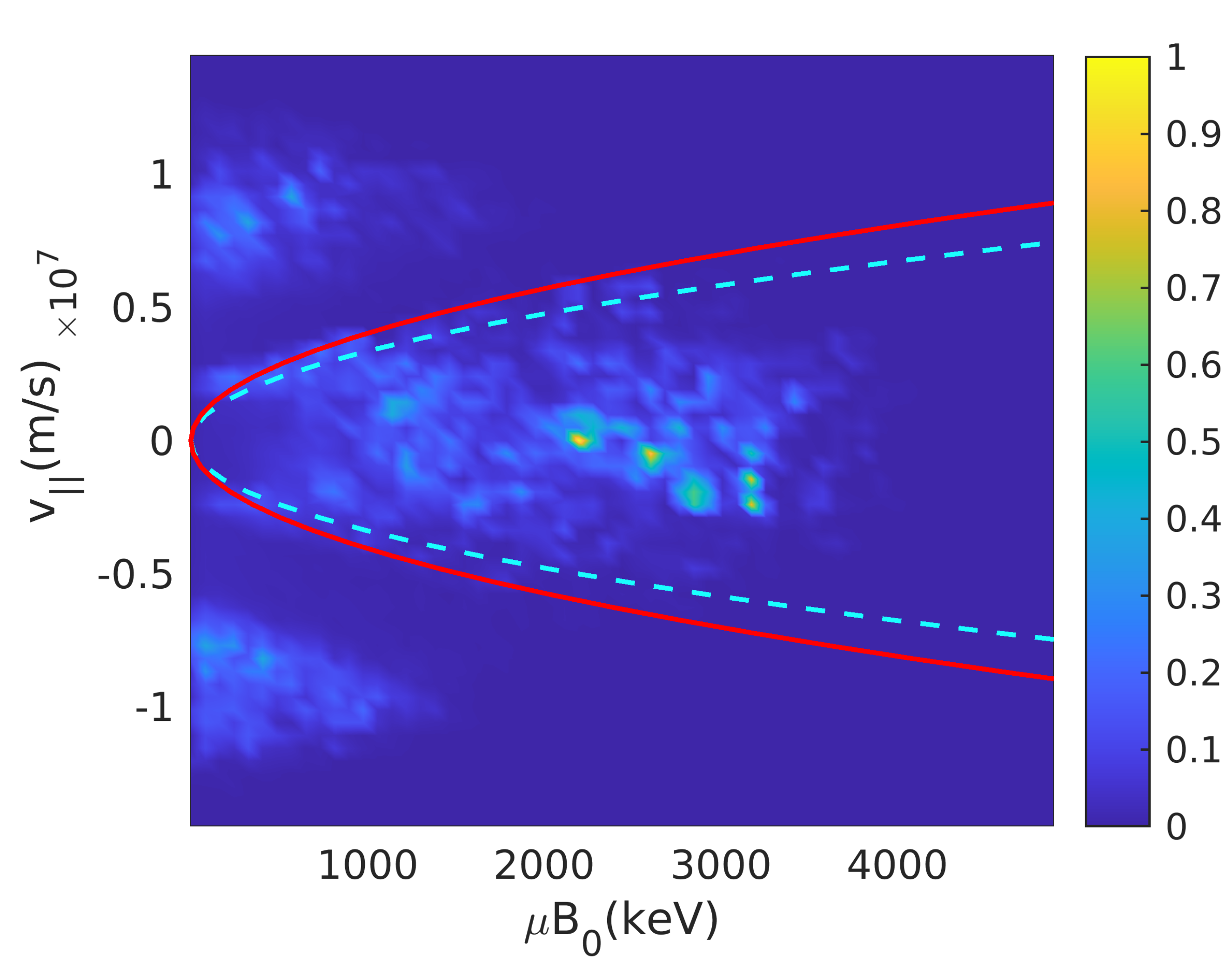}
\caption{(Color online) Normalized energy exchange in arbitrary units between EP and the mode integrated over a radial shell around the mode localization ($0.36<s<0.51$). The dashed cyan and solid red lines approximately indicate the trapped-passing boundary for the inner and outer radial surface of the considered shell, respectively.}\label{fig:Enexg}
\end{figure}
\begin{figure}[htbp]\centering
\includegraphics[width=0.4\textwidth]{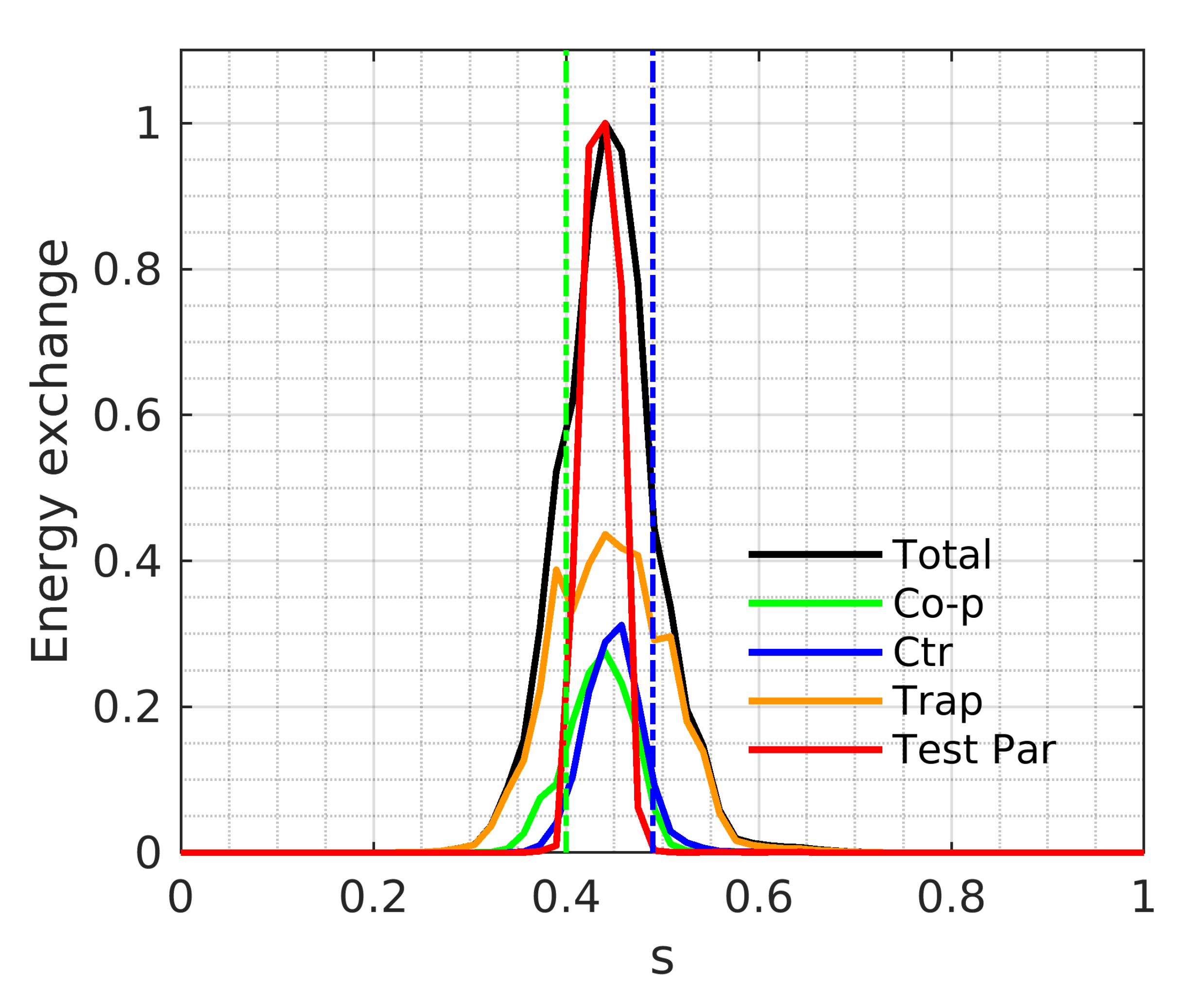}
\caption{(Color online) Radial profiles (normalized to one) of energy exchange contributed by different populations: Co-p, Ctr and Trap particles are separated and represented in green, blue and orange lines, respectively. The black line corresponds to the total energy exchange of all co/counter passing and trapped particles. The red line is the normalized radial profile of the test particles with $K=1.5$MeV and $\mu B_0 = 164$keV.	}\label{fig:Enexg_radial}
\end{figure}
  

The description of the resonant particles is the key issue of EP transport modeling. Here, we define the total energy exchange of particles interacting with modes as
\begin{align}
\Delta\mathcal{E}=-\sum_j\int_0^t \dot E_j \cdot \delta n_j dt\;,
\end{align}
where the time derivative of particle energy results to be $\dot E_j=d\mathcal{H}_j/dt=-\partial \mathcal{L}_{int}/\partial t$ (here $\mathcal{L}_{int}$ denotes the wave-particle interaction Lagrangian). This corresponds to the only energy transfer channel in HAGIS and the mode structure is fixed \cite{pinches1998hagis}. The weighting factor of markers is $\delta n_j=\delta f_j\Delta \Theta^{(p)}_j$, where $\Delta \Theta^{(p)}_j$ is the finite physical phase-space volume element associated with each marker.

In \figref{fig:Enexg}, we show the energy exchange (normalized to the maximum in arbitrary units) from the initial stage until saturation in the $(v_\parallel,\mu)$ phase space integrated over a radial shell around the mode localization ($0.36<s<0.51$). 
The energy exchange is contributed mainly by the most resonant particles. The contribution from co-passing (Co-p), counter-passing (Ctr) and trapped (Trap) particles clearly emerges from the plot. The resonance structure of passing particles (both Co-p and Ctr) in the $(v_\parallel,\mu)$ space is narrower. This energy exchange can be also integrated over $v_\parallel$ outlining a peak at $\mu B_0=164$\,keV corresponding to the passing resonance. 
The radial profiles of energy exchange from different populations are shown in \figref{fig:Enexg_radial}. The energy exchange of Co-p, Ctr and Trap particles are separated and depicted in green, blue and orange lines, respectively. The value is normalized to the maximum of total particles. 
Summing the energy exchange from different populations, the fraction of total energy exchange from the corresponding orbit types is evaluated as: Trap = 0.59; Co-p=0.2014; Ctr=0.2.

In order to get information about the resonant particle behaviour and its connection with mode saturation, we follow the evolution of a set of test particles (resonant particle sample) in the ﬁelds computed from the self-consistent simulation. These test particles do not contribute to the mode growth but serve as a diagnostic (details can be found in \citer{BW14pop}). As a test case to analyze, we take into account only the Co-p tracers: the slice which maximize the energy exchange can be set by fixing $K=1.5$ MeV and $\mu B_0=164$ keV (see \figref{fig:Enexg_radial}, red line). The evolution of this single slice can be now compared to the mapped analogous of the 1D reduced model. Since we have evaluated that the Co-p population contributes to the driving mechanism with a fraction $x=0.2014$, we can properly consider that the dynamics of the addressed slice is characterized by $\bar{\Gamma}^*/\bar{\Omega}=x\, \bar{\Gamma}/\bar{\Omega}$. As already introduced, we now verify that the reduced model can reproduce the slice transport without using the scale factor in the drive relation. In this sense, we apply the mapping procedure by setting 
\begin{align}
\bar{\gamma}_D/\bar{\omega}=\bar{\Gamma}_D^*/\bar{\Omega}\;,
\end{align}
instead of \eref{scaling}. We recall that we use this drive relation to integrate the linear dispersion relation \eref{disrel} obtaining the parameter $\eta$ for running the simulation of the BoT model and reproducing the radial profile via the mapping $u\to s$. As initial particle distribution function we use the initial tracer profile used in the analysis above.
\begin{figure}[ht!]
\centering
\includegraphics[width=.4\textwidth,clip]{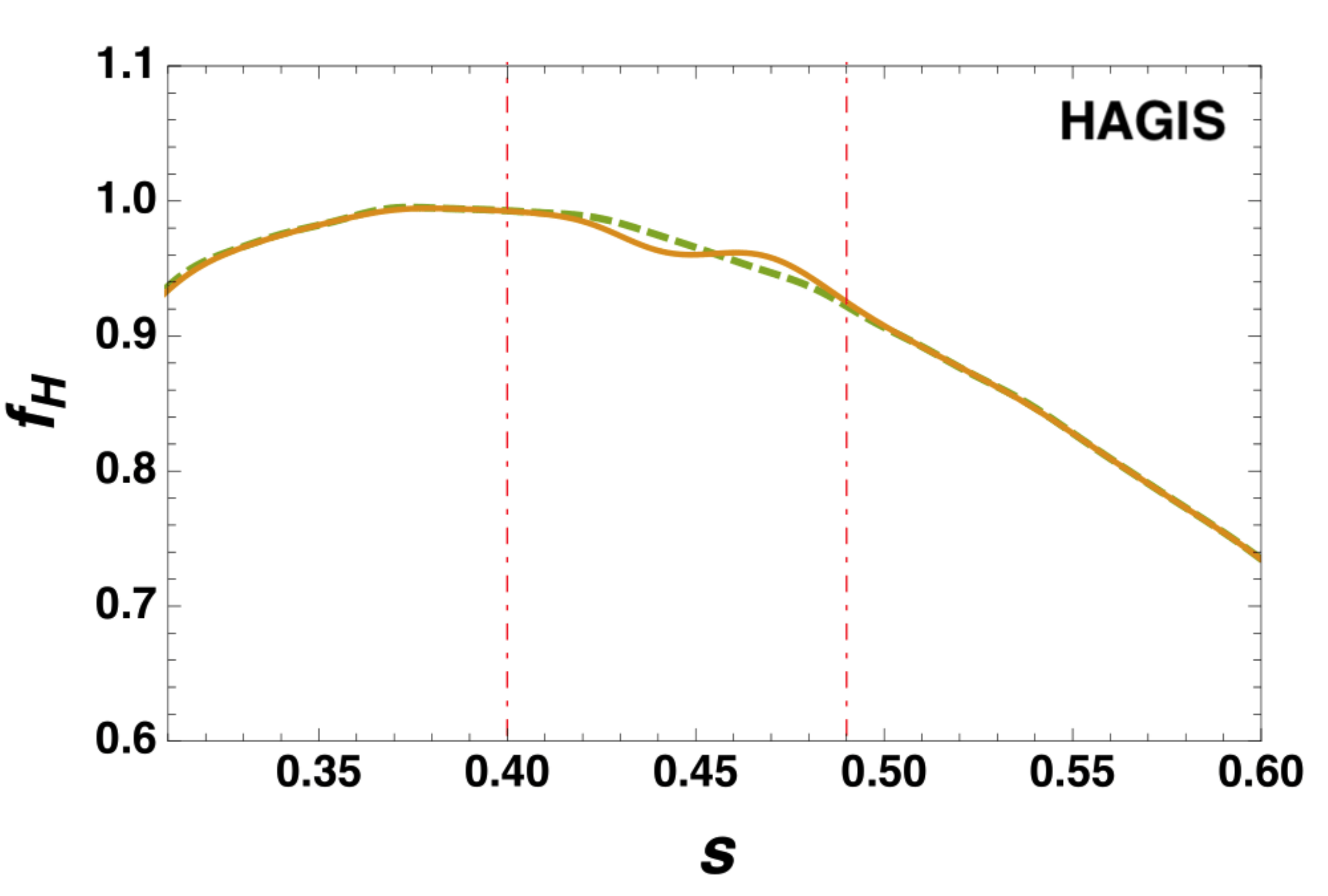}\\
\includegraphics[width=.4\textwidth,clip]{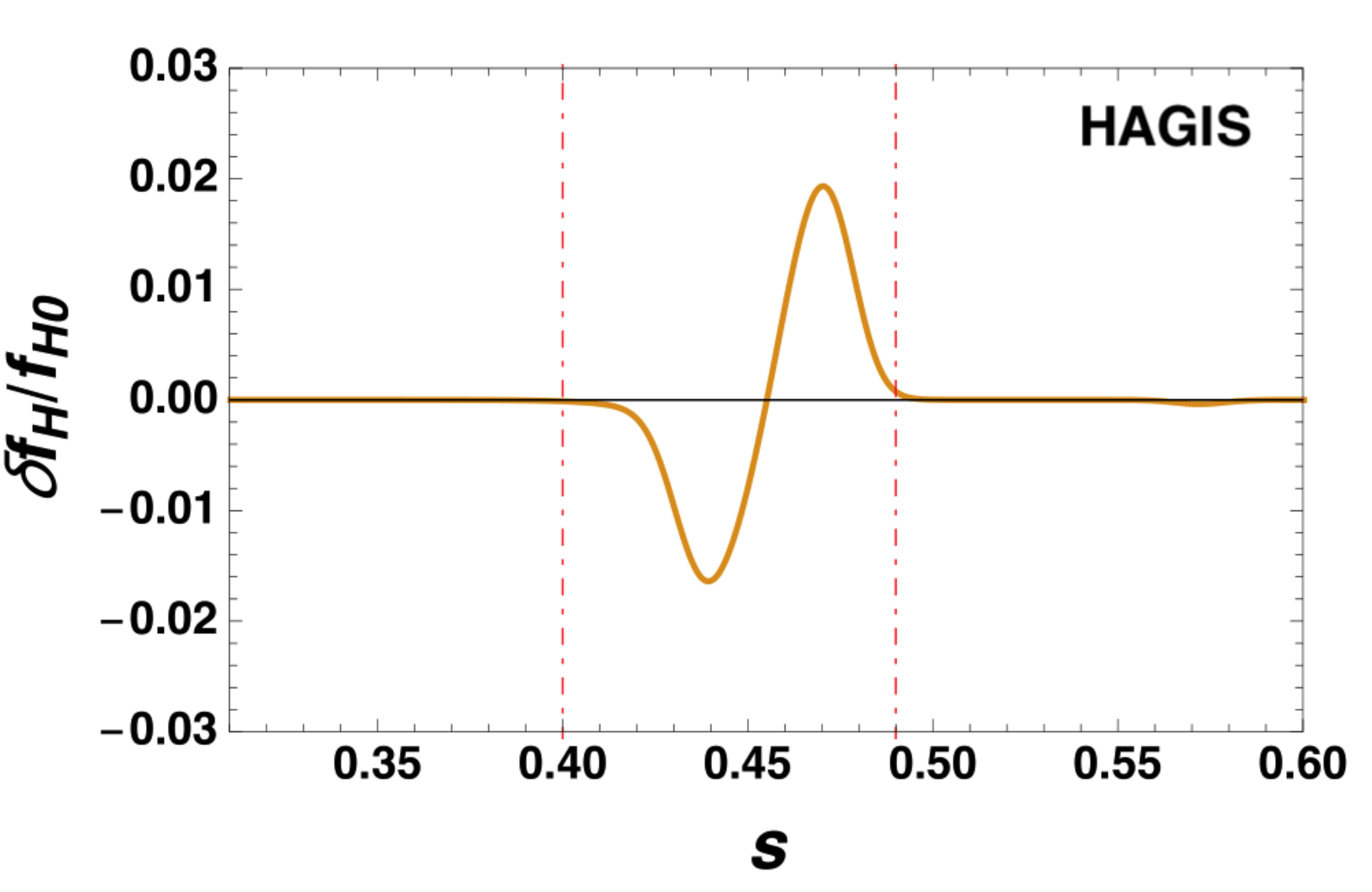}
\caption{(Color online) Upper panel - Redistribution at saturation of the most resonant particle from HAGIS simulations (dashed green line is the initial distribution). Lower panel - Detail of the profile perturbations with respect to the initial values. Vertical red lines define the flattening region width.
\label{fighagis}}
\end{figure}
\begin{figure}[ht!]
\centering
\includegraphics[width=.4\textwidth,clip]{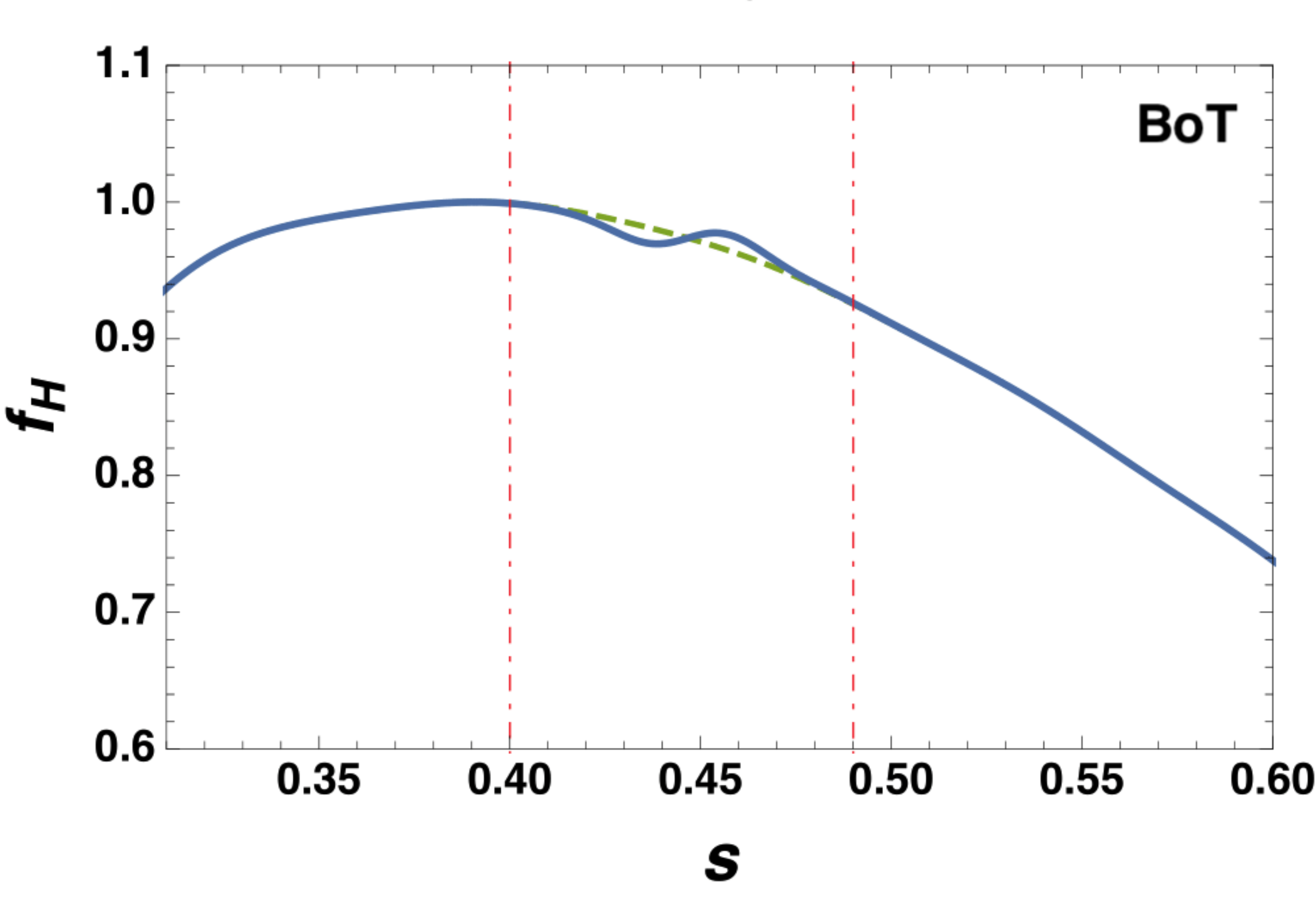}\\
\includegraphics[width=.4\textwidth,clip]{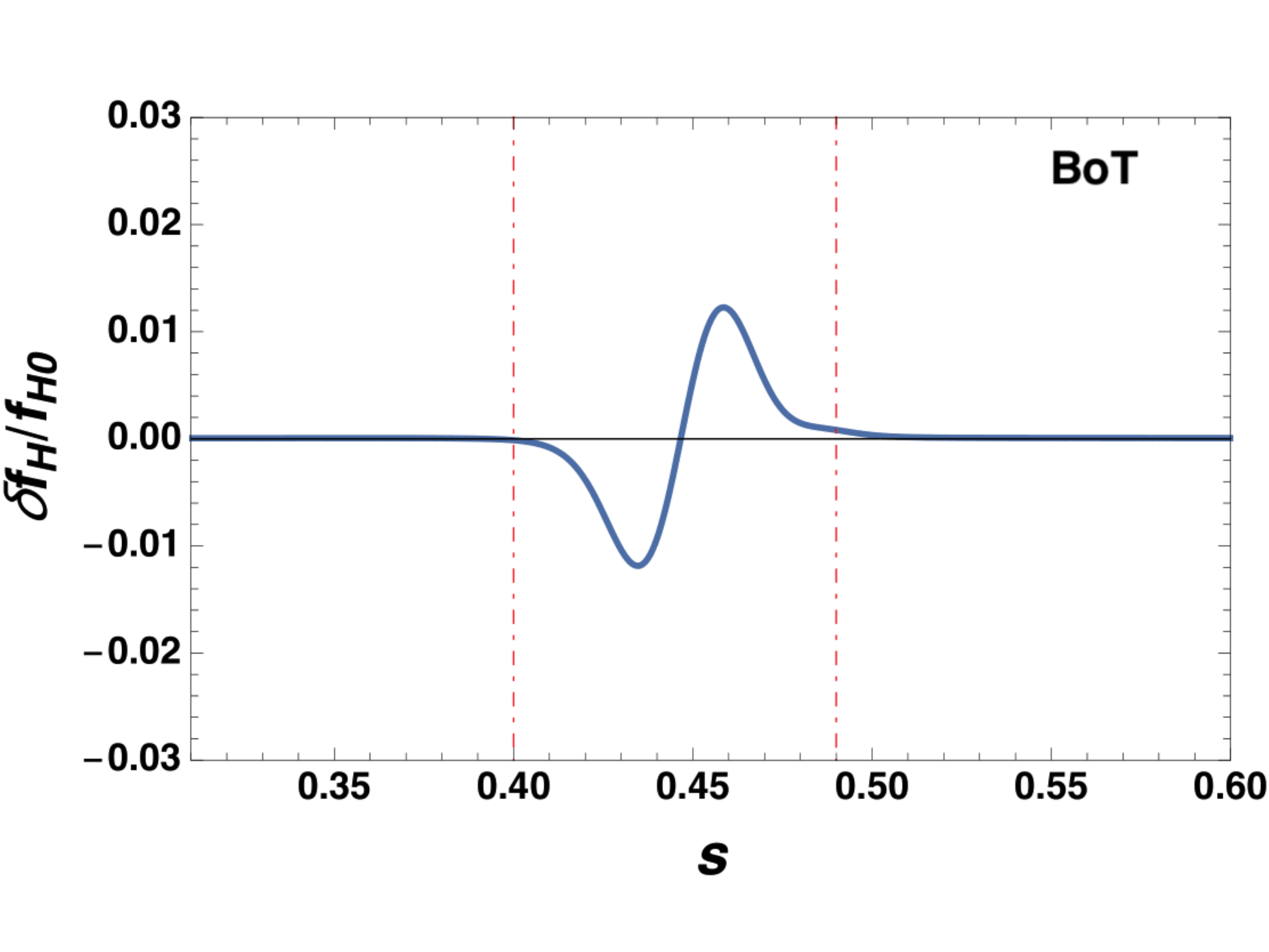}
\caption{(Color online) Reduced model prediction of the dynamics in \figref{fighagis}.
\label{fighagisbps}}
\end{figure}

In \figref{fighagis}, the evolution of the single slice profile at saturation is shown. It emerges how the flattening is now clearly evident with respect to the almost unperturbed global radial profile in \figref{fig:del_f}. The mapped evolution of the most resonant slice is instead plot in \figref{fighagisbps} outlining the very good predictive capability of the reduced model.

The analysis of this Section shows the capability of the mapping procedure and of the 1D reduced model to properly describe the redistribution of the most resonant particles, \ie the EP belonging to a defined slice in the phase space which maximizes the wave particle power exchange. In this sense, the drive reduction implemented in \eref{scaling} is mandatory only when the global dynamics of EP is taken into account. This demonstrates that the crucial problem of the dimensional reduction when mapping the realistic scenario with the reduced 1D model is actually due to the proper representation of non-resonant EPs. 

We want to stress how this analysis can be seen as a starting point for further development beyond the scope of the present work. In fact, by cutting the phase space into a sufficient number of distinct slices (representing the global wave-particle energy exchange), the reduced model can be evolved, in parallel, with different set up representing each single slices. Then, the dynamics could be summed up consistently resulting into the global EP redistribution.

\section{Concluding remarks}
We constructed a mapping between the velocity space of a BoT model, described via an Hamiltonian $N$-body approach, and the radial coordinate related to the transport properties of fast ions interacting with Alfv\'enic fluctuations in a Tokamak device. In agreement with the ideas developed in \citers{BB90a,BB90b,BB90c}, we propose and demonstrate the existence of a direct relation between the transport features in the two different contexts on the basis of a correspondence of the drive mechanism. That is, the gradients in the velocity space responsible for the inverse Landau damping are equivalent to the radial gradients of the pressure in the toroidal plasma. The possibility to reduce the transport features of fast ions in Tokamaks to an essentially 1D problem has been successfully pursued in \citer{GG12}, starting from the QL two-dimensional model derived in \citer{K72}, but the predictivity of this analysis is limited to the diffusive radial behavior of the fast particles.

In \citer{spb16}, ITER relevant simulations have been performed, outlining deviations from the QL diffusion of the radial EP transport. In fact, the possibility for avalanche processes has been clearly inferred, mainly dominated by convection rather than diffusive transport. Our study has the merit to demonstrate how such new features can be predicted by defining a proper mapping procedure, which can link the BoT paradigm to the initial conditions of the numerical simulations in \citer{spb16}. Important to note, here, is that the mapping between the realistic 3D system and the 1D BoT reduced model can be constructed, \emph{de facto}, by means of linear numerical simulation results. The predictivity of the reduced description, meanwhile, is extended to nonlinear EP transport and is inclusive of the description of EP avalanches and convective transport. This is what makes this reduced model approach particularly attractive and motivates its further extension to improve its predictive capability to multiple resonances.

Our analysis successfully demonstrates that the deviation from the QL evolution, observed in the ITER relevant analysis, can be properly interpreted as avalanche phenomena. 
As a consequence, the spectrum and distribution function evolution are deformed with respect to a QL prediction: the fattening of the distribution function is altered and evidence for the formation of traveling local clumps emerges.

The capability of the reduced 1D model in reproducing the relevant physical features of the particle-wave interaction must be considered together with the important computational saving (10000 time less expensive) of the $N$-body code. This two features constitute the power of the addressed procedure and are at the ground of the practical applicability of the 1D model. In particular, various scans of linear growth rates and/or the amount of resonance overlap can be performed implementing fast non-linear simulations of the reduced scheme, discriminating the different transport regimes of specific cases and providing information regarding the necessity of a fully 3D expensive run.

Despite the successes of our mapping predictions, at the present level, the agreement with the LIGKA/HAGIS simulations still remains on a qualitative level. This is due to many subtleties of the constructed paradigm and below we report the most relevant ones, which require attention before a satisfactory quantitative prediction of the fast ion fluxes in Tokamaks is achieved:

\begin{enumerate}
\item[(a)] Here, we adopt a linear map between velocities and radial positions. Actually, this is a solid assumption as far as the radial change of the resonant frequency  $\Omega_F$ is smooth and we can retain only linear terms in the Taylor expansion. When we consider resonances which are fairly distant in the $r$-space from the one chosen to construct the mapping, some deviation from the linearity can emerge in the expansion, affecting the scaling of the basic model parameters. This suggests a more careful analysis of the radial behavior of the resonant frequency and, in principle, the construction of a nonlinear or continuum mapping, able to precisely reproduce the radial structures of the resonances.

\item[(b)] The possibility to infer the radial non-linear spread of EP, from the one observed in the velocity space, \ie the proper definition of the mode growth rates, is strongly limited by the difference existing in the phenomenology of the two systems and therefore it require a suitable calibration, before the mapping can be correctly applied. 
As discussed above, this is mostly due to the different fraction of resonant particles in the physical system and the 1D reduced model. Suggestions about how this shortcoming can be removed are given in the text.

\item[(c)] The analysis in \citer{spb16} is performed evolving all the fast particles simultaneously, disregarding, as it must be, if they belong to different slices defined by the adiabatic constants of motion. This is a typical 3D feature of the model, but, clearly, particles from different slices (corresponding to different first integrals) exchange energy with the fields on a different level. This aspects is here analyzed for a single resonance and will be globally addressed in the mapping in future works, similar to the mitigation strategy for item above.

\item[(d)] Our analysis is possible as far as the radial mode structure is large with respect the non-linear radial excursion of the EP orbits. In fact, there is no way in the BoT paradigm to account for a mode structure in the velocity space, since the mode amplitude is constant over the velocity space for a given wave-number. This is probably the easiest issue to correct, since the failure of the mapping procedure is expected in such a situation.

\end{enumerate}

Despite all these shortcomings of the mapping procedure, the present study demonstrates its validity for an estimation of what is the radial transport of fast ions in an ITER configuration. Further refinements of this paradigm, in the spirit traced by the points above, are reliably able to transform it into a quantitative reduced model to predict the real EP fluxes in realistic Tokamak geometry.

\vspace{0.5cm}

{\small ** This work has been carried out within the framework of the EUROfusion Consortium [Enabling Research Projects: MET (CfP-AWP19-ENR-01-ENEA-05) and ATEP (CfP-FSD-AWP21-ENR-03)] and has received funding from the Euratom research and training programme 2019-2020 under grant agreement No 633053. The views and opinions expressed herein do not necessarily reflect those of the European Commission. **}

\vspace{1cm}


\end{document}